\documentclass[aps,prb,twocolumn,showpacs,final,floatfix,longbibliography,superscriptaddress]{revtex4-2}
\usepackage{amsmath}
\usepackage{amsfonts}
\usepackage{amssymb}
\usepackage{physics}
\usepackage{color}
\usepackage{hyperref}
\usepackage{inputenc}
\usepackage{enumerate}
\usepackage{graphicx}
\usepackage{float}
\usepackage{amsmath}
\usepackage{amsfonts}
\usepackage{lipsum}
\usepackage{dcolumn}
\usepackage{hyperref}
\usepackage{subfigure}
\usepackage{epsfig}
\usepackage{epsf} 
\usepackage{epstopdf}

\usepackage{color}
\usepackage{enumitem}
\usepackage{xr}
\makeatletter

\newcommand*{\addFileDependency}[1]{
\typeout{(#1)}
\@addtofilelist{#1}
\IfFileExists{#1}{}{\typeout{No file #1.}}
}\makeatother

\usepackage{hyperref}
\hypersetup{colorlinks,
    citecolor=blue,
    filecolor=cyan,
    linkcolor=blue,
   urlcolor=magenta
 }

\usepackage{cleveref}

\usepackage{hyperref}
\hypersetup{
    colorlinks,
    citecolor=red,
    filecolor=cyan,
    linkcolor=blue,
   urlcolor=magenta
 }

\begin{document}

\title{Yang-Lee Zeros of 2D Nearest-Neighbor Antiferromagnetic Ising\\ Models: A Numerical Linked Cluster Expansion Study}


\author{Mahmoud Abdelshafy}
\altaffiliation{These authors contributed equally to this work}
\affiliation{Department of Physics, The Pennsylvania State University, University Park, Pennsylvania 16802, USA}

\author{Muhammad Sedik}
\altaffiliation{These authors contributed equally to this work}
\affiliation{Physics Department, University of California, Santa Cruz, California 95064, USA}




\date{\today}

\begin{abstract}
We study Yang-Lee zeros as they approach the thermodynamic limit of the 2D nearest-neighbor antiferromagnetic Ising model on square and triangular lattices. We employ the Numerical Linked Cluster Expansion (NLCE) equipped with Exact Enumeration (EE) of the partition function to estimate the Laplacian of the free energy, which is proportional to the zeros density. Using a modified NLCE, where the expansion can be carried directly on the Yang-Lee zeros of the involved clusters, we estimate the density of Yang-Lee zeros in the thermodynamic limit. NLCE gives significantly more zeros than EE in the complex field plane providing more insights on how the root curves look in the thermodynamic limit. For the square lattice at $T \ll T_c$, the results suggest that two vertical lines at $\pm h_c(T)$ in the complex field plane (i.e two concentric circles in the complex fugacity plane) are the thermodynamic root curves. A similar picture is found for the triangular lattice for phase transitions at large values of magnetic field while further study is needed for phase transitions at smaller values of magnetic field. The convergence of the NLCE and (EE) calculations of the partition function to the thermodynamic limit is studied in both lattices and the temperature-field phase diagram is obtained from Yang-Lee zeros using both methods. This NLCE-based approach will facilitate the study of different types of phase transitions using Yang-Lee zeros in future research.
\end{abstract}

\maketitle

\section{Introduction}
The study of phase transitions plays a central role in condensed matter physics leading to some of the most celebrated advancements in the field, e.g. criticality and scaling theories and the renormalization group approach~\cite{pathria2011,domb_green_book_74}. A phase transition describes a substantial change to the steady state of a physical system at thermal equilibrium as a control parameter is varied. Classic examples include liquid-gas transitions, and the paramagnet-(anti)ferromagnet transitions in spin systems. To study phase transitions quantitatively, one typically introduces a thermodynamic potential, e.g. free energy, as a function of its control parameters. The existence of a phase transition is usually manifested in the non-analytic behavior of the thermodynamic potential at the transition point, namely the critical point. Since thermodynamic potentials are proportional to the logarithm of the partition function, phase transitions should occur at points where the partition function diverges or vanishes.

Motivated by this picture, Yang and Lee parametrized the partition function of the Ising model as a polynomial\cite{yang1952statistical}. Hence, only the zeros of the partition function can lead to non-analytic behavior in the free energy and they proposed to study the distribution of the partition function zeros in the complex field plane as a means to identify phase transition points for physical systems~\cite{yang1952statistical}. In the thermodynamic limit, the number of zeros increases approaching a continuous limit, and forming whether dense areas \cite{van1984location,mussardo2017yang,bencs2022lee} or root curves \cite{lee1952statistical,hemmer1964yang,sedik2024yang}. These root curves intersect with the real axis only at critical points. For Ising-type models, the focus of our study, Lee and Yang proved a phase transition can only exist at zero external field for ferromagnetic interactions with any interaction range, dimensionality, and geometry in their famous Lee-Yang circle theorem~\cite{lee1952statistical}. The scope of applications of the Yang-Lee approach has been expanded over the years to include other classical~\cite{katsura1954phase,heilmann1971location,newman1974zeros,lieb1981general,biskup2000general,deger2020lee,li2023lee} and quantum equilibrium phase transitions~\cite{asano1970theorems,vecsei2022lee,li2023yang,li2024yang}, among others~\cite{bena2005statistical}. Moreover, the Yang-Lee zeros distribution has garnered increasing interest in experimental realizations in recent years~\cite{peng2015experimental,brandner2017experimental,gao2024experimental,brange2024lee}. 

One area that still lacks conceptual understanding is Ising-type models with antiferromagnetic interactions. Analytical approaches have been restricted to either the high temperature regime ~\cite{heilmann1970monomers,lieb1972property,sedik2024yang} or certain geometries~\cite{lebowitz2012location}, while numerical studies using exact enumeration (EE) are limited due to the finite-size effects in addition to the sparsity of the zeros, which prevents one from determining their loci in the thermodynamic limit~\cite{kim2004yang,hwang2010yang,sedik2024yang}. Although Monte-Carlo techniques allow to explore larger system sizes, they suffer from instabilities in the precision when finding all the zeros~\cite{sedik2024yang}. To overcome these adversities, one can opt for a "free energy" approach in which finding the free energy per site in the thermodynamic limit immediately gives the density of Yang-Lee zeros in the thermodynamic limit~\cite{bena2005statistical,hemmer1964yang,sedik2024yang}. However, for nearest-neighbor Ising models, finding the thermodynamic free energy per site with high enough precision for complex-valued magnetic fields is still a challenging task due to the aforementioned reasons.

An intriguing technique to capture the thermodynamic limit behavior of lattice models is the numerical linked cluster expansion (NLCE)~\cite{rigol_bryant_06,rigol_bryant_07a,rigol_bryant_07b}. NLCEs can be used to obtain extensive thermodynamic observables directly in the thermodynamic limit for spin lattice models~\cite{Tang_2013}. NLCEs have a number of advantages including: (1) The freedom to choose different building blocks for a given lattice where various choices proved efficient for different applications~\cite{rigol_bryant_07a,richter_20,Abdelshafy_23,Abdelshafy_24}, (2) The absence of any expansion parameter making the series potentially convergent as long as the correlations do not grow beyond the largest clusters included in the expansion~\cite{iyer_15,Abdelshafy_23}, and (3) the flexibility to have other numerical techniques implemented in combination with NLCEs to take advantages of the combined techniques~\cite{bruognolo2017matrix,khatami2014linked,hagymasi2021possible,biella2018linked,richter2019combining}. 

In this work, we combine the NLCE with the free energy approach to study the thermodynamic limit distribution of the Yang-Lee zeros for the nearest-neighbor antiferromagnetic Ising model on the square and triangular lattices. We achieve this by picking the rectangle expansion~\cite{gan_20,richter_20,Abdelshafy_24} as the NLCE of choice and use Bhanot's Algorithm~\cite{bhanot1990numerical,creswick1995transfer}, extended to include a magnetic field~\cite{sedik2024yang}, to calculate the zeros of the partition function for clusters as large as 256 sites. We introduce a modified version of the linked cluster expansion that performs the expansion directly on the Yang-Lee zeros of the involved clusters. This allows us to gain more insights on how the root curves look like as they approach the thermodynamic limit for the considered models. Using the largest NLCE order obtained, we show the regions in the complex field plane where the zeros intersect with the real axis in the thermodynamic limit signaling phase transitions. We conclude by depicting the temperature-field phase diagram for both lattice geometries utilizing the extrapolation done from the NLCE results and comparing it to phase diagrams obtained in previous studies. 

The paper is organized as follows. In Sec.~\ref{Sec:YL-zeros}, we introduce the model and the definition of the Yang-Lee zeros. We discuss the free energy approach in Sec.~\ref{sec:free-energy-approach}. Then, we briefly cover the NLCE and the modifications we introduce to apply it directly on the level of the Yang-Lee zeros in Sec.~\ref{Sec:NLCE}. In Sec.~\ref{SubSec:square}, we study the Yang-Lee zeros distribution in the thermodynamic limit for the nearest-neighbor Ising model on a square lattice. We benchmark our NLCE method for the case of ferromagnetic interactions, and then investigate the zeros distribution for the less-explored antiferromagnetic version of the model. We carry out a similar investigation for the triangular lattice in Sec.~\ref{SubSec:triangular}. We conclude in Sec.~\ref{Sec:Summary} with a summary and discussion.

\section{Model and Yang-Lee Zeros}\label{Sec:YL-zeros}
We study the classical Ising model with an external field written in the form
\begin{equation}
     \mathcal{H}=J\sum_{\langle ij \rangle}(1-\sigma_i\sigma_j)-h\sum_{i}(1+\sigma_{i}),
    \label{H1}
\end{equation}
where $\sigma_i=\pm 1$ denotes a spin at site $i$ on a square or a triangular lattice geometry with a total number of sites $N_s$ and a total number of bonds $N_b$ depending on the boundary conditions. $J$ is the homogeneous interaction strength between nearest neighbor spins, denoted by $\langle ij \rangle$, and $h$ is an applied external magnetic field. The partition function then follows as a sum over all possible spin configurations $\{\boldsymbol{\sigma}\}$
\begin{equation}
        \mathcal{Z}_{N_s}=\sum_{\{\boldsymbol{\sigma}\}}e^{-\beta \mathcal{H}}=\sum_{M=0}^{N_s}\sum_{E=0}^{N_b}\Omega(E,M)e^{-2\beta JE}z^{M}.
        \label{eqn:Z(omega)}
\end{equation}
Here, $z=e^{2\beta h}$ and $\beta= \frac{1}{k_B T}$, where $k_{B}$ is the Boltzmann constant. We set $J$ and $k_B$ to unity for the rest of this work. $\Omega(E,M)$ counts the number of states with integer interaction energy $E$ and integer magnetization $M$, which are given by
\begin{equation}
        E=\frac{1}{2}\sum_{\langle ij \rangle}(1-\sigma_i\sigma_j),  ~ M=\frac{1}{2}\sum_{i}(1+\sigma_i).
    \label{IntEnergy/Mag}
\end{equation}
We define $a_{M}=\sum_{E=0}^{N_b}\Omega(E,M)e^{-2\beta E}$ to write the partition function in a compact polynomial form $ \mathcal{Z}_{N_s}=\sum_{M=0}^{N_s}a_{M}z^{M}$, which will prove convenient for the subsequent analysis. Therefore, at a fixed temperature, this monic polynomial partition function in the variable $z$ can be uniquely recast in the formula
\begin{align}
\mathcal{Z}_{N_s}=\prod_{j=1}^{N_s}(z-z_j),
\end{align}
where $z_j$ are the zeros of the partition function, dubbed the Yang-Lee zeros. By the fundamental theorem of algebra, a monic polynomial is determined uniquely by its roots; hence, the thermodynamic state of a given system can be fully identified by knowledge of its Yang-Lee zeros. Thus, studying the distribution of the Yang-Lee zeros in the complex $z$ plane, i.e. the complex fugacity plane, can lead to the identification of phase transitions, if they exist for a given system, as their distribution would touch the positive real axis at the critical values of fugacity~\cite{lee1952statistical}. If the Yang-Lee zeros touch the real axis in the thermodynamic limit, this would indicate a phase transition because of the non-analyticity of the free energy per site at these points, $\beta \Tilde{f}=\lim_{N_s\rightarrow\infty}(-1/N_s)\ln(\mathcal{Z})$. 

The Ising Hamiltonian is invariant under the $h\rightarrow -h$ symmetry, which is reflected in $z_{j} \rightarrow 1/z_{j}$. Hence, one needs only to focus on studying the roots within the unit disk $|z_{j}|\leq 1$. Moreover, if $z_{j}$ is a root of the partition function, then as a result, $z_{j}^*$ is also a root because the coefficients $a_M$ are real. When studying the Yang-Lee zeros distribution in the $h$ plane, where $h_j=\frac{1}{2\beta}\ln(z_j)$, this would be reflected in observing $h_{j} \leftrightarrow h_{j}^*$ as well.

\section{Free Energy Approach}
\label{sec:free-energy-approach}
The intensive free energy $\Tilde{f}$ is given by $\Tilde{F}/N_s=-\ln(\mathcal{Z}_{N_s})/(\beta N_s)$, where $\Tilde{F}$ is the extensive free energy. The real part of the intensive free energy can be expressed in terms of the Yang-Lee zeros as follows:
\begin{equation}
    \beta f(z) = \beta \Re(\Tilde{f}(z))=\frac{1}{N_s}\sum_{j=1}^{N_s}\ln|z-z_j|.
    \label{FiniteReFE}
\end{equation}
An intuition of this expression of the free energy can be developed in analogy with elementary electrostatics. The real part of the free energy can be interpreted as the potential due to $N_s$ point charges located at points $z_j$ on a two-dimensional plane. Therefore, the partition function roots can be determined by calculating the Laplacian of the real part of the free energy per site. In fact, in the thermodynamic limit, this Laplacian would directly give the density of the roots, $\rho(z)$. We can therefore write, 
\begin{equation}
    \rho(z)= -\frac{1}{2\pi}\nabla^2\left[\beta f(z)\right].
    \label{PoissonEq}
\end{equation}
Since $\rho(z)$ and $f(z)$ are scalar quantities, Eq.~(\ref{PoissonEq}) holds under coordinate transformation from $z$ to the complex $h$ plane. A similar approach of calculating the density of Lee-Yang zeros was explained in Ref.~\cite{bena2005statistical}. Hemmer and Hauge used this technique to study the Yang-Lee zeros of a van der Waals gas analytically through the equation of state~\cite{hemmer1964yang}. In addition, one of the authors used it to find new root curves in the thermodynamic limit of the mean field model through the self consistency equations~\cite{sedik2024yang}. For models where such an analytic approach is still intractable, an approximation to the thermodynamic free energy should give an approximation to the density of the zeros in the thermodynamic limit. In the next section, we show how the density of Yang-Lee zeros in the thermodynamic limit could be approximated by means of the NLCE employing Eq.~(\ref{PoissonEq}). We then introduce the particular expansions we use for the lattices of interest here, namely the square and the triangular lattices. 

\section{NLCE}\label{Sec:NLCE}
NLCE is a series expansion technique that can be used to calculate extensive observables per lattice site $\mathcal{O}/N_s$ for arbitrary translationally invariant lattices. The idea is to sum over contributions from all connected clusters --based on a consistent criterion that we will specify later-- that can be embedded on the original lattice as follows
\begin{equation}\label{eq:nlce}
\mathcal{O}/N_s=\sum_{c} L(c)\times W_{\mathcal{O}}(c).
\end{equation}
$W_{\mathcal{O}}(c)$ is the weight of observable $\mathcal{O}$ in a connected cluster $c$, and $L(c)$ is a geometric factor counting the number of ways per site this cluster can be embedded on the original lattice. We refer the reader to Ref.~\cite{Tang_2013} for a comprehensive introduction to NLCEs. The weights in Eq.~\eqref{eq:nlce} are calculated through the recursive relation:
\begin{equation}\label{eq:weight}
W_{\mathcal{O}}(c)=\mathcal{O}(c)- \sum_{s\subset c} W_{\mathcal{O}}(s),
\end{equation}
where we set $W_{\mathcal{O}}(c)=\mathcal{O}(c)$ for the smallest cluster considered. Typically, the observable $\mathcal{O}(c)$ for cluster $c$ is calculated at thermal equilibrium at temperature $T$ through a grand-canonical ensemble average (with zero chemical potential):
\begin{equation}
\mathcal{O}(c)=\frac{\sum_{i} o_{i}(c)e^{-\beta \epsilon_{i}(c)}}{\mathcal{Z}_{N_s}(c)}, \;\; \text{where} \;\; \mathcal{Z}_{N_s}(c)=\sum_{i} e^{-\beta \epsilon_{i}(c)}.
\end{equation}
Here, the sums run over all the microstates with energy $\epsilon_i$ and $\mathcal{O}=o_i$ for cluster $c$ with $N_s$ sites. To approximate observables in the thermodynamic limit, i.e. for infinitely large lattices $N_s \rightarrow \infty$, one truncates the sum in Eq.~(\ref{eq:nlce}) at some finite order --usually set by computational limitations. The orders in Eq.~(\ref{eq:nlce}) are conventionally labeled by the number of sites of the largest cluster considered in a given order, and the series is considered convergent at a given temperature $T$ when the results of successive orders agree with each other to machine precision. Normally, NLCEs provide precise results as long as correlations between spins do not grow beyond the size of the largest cluster considered in the sum. This growth results in a failure of NLCE near critical temperatures where models develop infinitely long correlations. Away from critical temperatures, NLCEs have been reported to exhibit exponential convergence to the thermodynamic limit results~\cite{iyer_15}.

As we mentioned before, in this work we aim to extend the applications of NLCEs to study the Yang-Lee zeros of the Ising model on square and triangular lattices. The observable of interest here is the real part of the intensive free energy $f$, which can be expressed in terms of the Yang-Lee zeros as indicated by Eq.~(\ref{FiniteReFE}). When we plug in the real part of the extensive free energy $F$ as the observable $\mathcal{O}$ into the NLCE equations, Eqs.~(\ref{eq:nlce}) and~(\ref{eq:weight}) read
\begin{eqnarray}\label{eq:freenlce}
&&f =\sum_{c} L(c)\times W_{F}(c),\nonumber \\
&&W_{F}(c)=-\frac{1}{\beta}\sum_{j=1}^{N_s(c)}\ln\left|z-z_j(c)\right|- \sum_{s\subset c} W_{F}(s),
\end{eqnarray}
where $N_s(c)$ denotes the number of sites in cluster $c$ and $z_j(c)$ are the Yang-Lee zeros of the partition function of cluster $c$. It becomes clear that knowledge of the Yang-Lee zeros of each cluster is enough to carry out the NLCE sum. When the series is truncated at a particular order $\ell$, the zeros of each cluster contribute to the sum with an integer factor $M(c)$ that depends on $L(c)$, the number of subtractions of $c$ as a sub-cluster from larger clusters, and the truncation order $\ell$. The NLCE sum could then be expressed as 
\begin{align}
    \beta f_{\ell}=-\sum_{c}M(c,\ell)\sum_{j=1}^{N_s(c)}\ln\left|z-z_j(c)\right|.
\end{align}

However, we do not need to calculate the bare free energy per site for our analysis, but rather its Laplacian, which is essentially a sum of Dirac-delta functions at the loci of the Yang-Lee zeros.
\begin{align}
    \rho_{\ell}(z)&=\frac{1}{2\pi}\nabla^2\left[\sum_{c}M(c,\ell)\sum_{j=1}^{N_s(c)}\ln\left|z-z_j(c)\right|\right] \notag\\
    &=\sum_{c}M(c,\ell)\sum_{j=1}^{N_s(c)}\delta^2\left[z-z_j(c)\right],
    \label{eqn:rho}
\end{align}
where $\delta^2\left[z-z_j(c)\right]=\delta\left\{\Re(z)-\Re\left[z_j(c)\right]\right\}\times$ \ $\delta\left\{\Im(z)-\Im\left[z_j(c)\right]\right\}$ is the two-dimensional Dirac-delta function in the complex $z$ plane. Thus, $\rho_{\ell}(z)$ could be used as an approximation to the thermodynamic limit density $\rho(z) = \lim_{\ell\rightarrow \infty} \rho_{\ell}(z)$. The Yang-Lee zeros result at a given NLCE order is simply all the zeros of all the clusters contributing to the sum, while the integer factor in front of each cluster in the sum represents an "amplitude" $M(c)$ of its corresponding zeros. Going back to the electrostatic analogy, the loci of the zeros are the loci of the charges and the factor in front of the zeros in the sum is the value of the charge at the corresponding location.
\begin{figure}[b]
   \centering
   \includegraphics[width=0.99 \columnwidth]{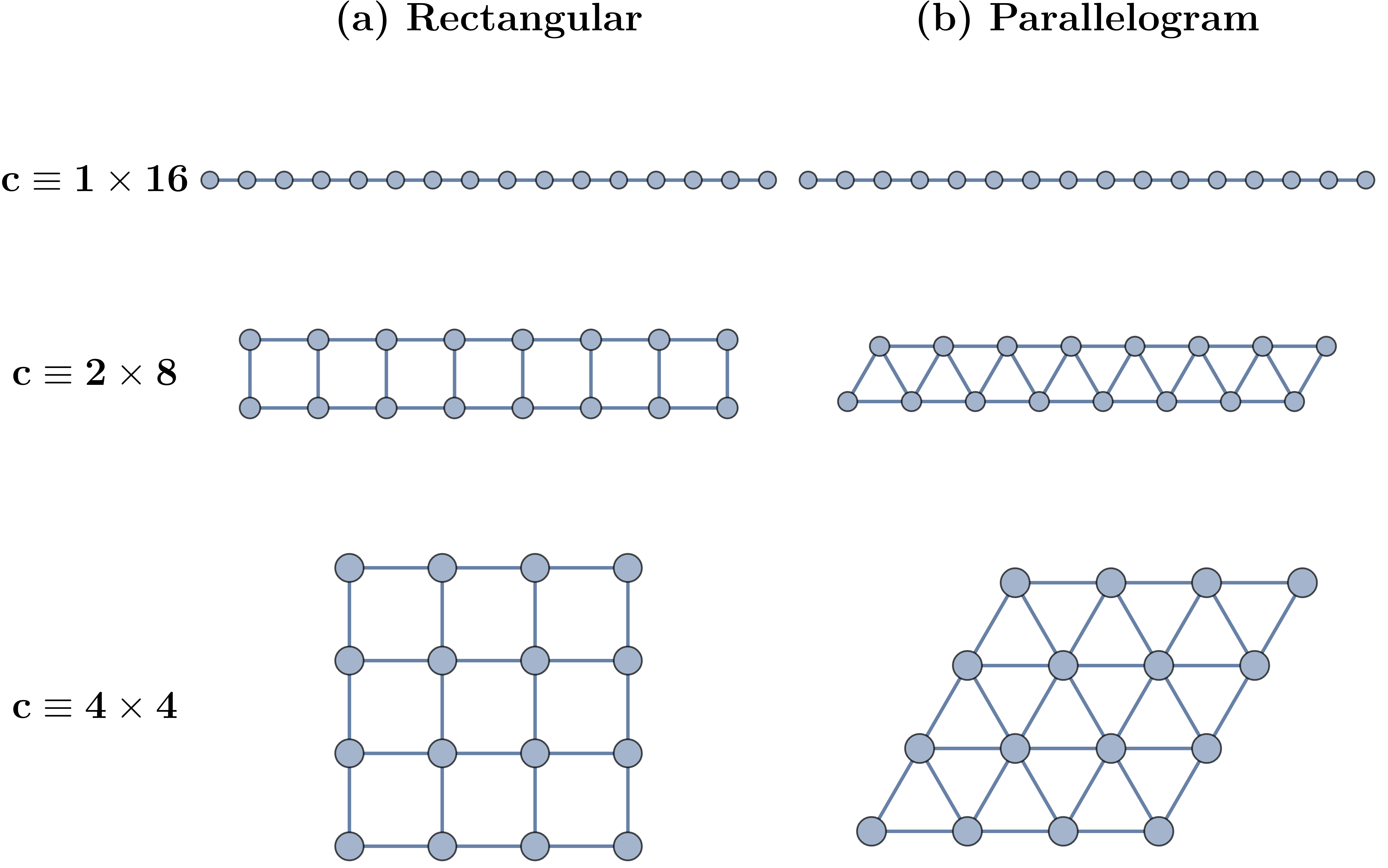}
   \caption{All possible (a) rectangular and (b) parallelogram clusters with $16$ sites.}
   \label{fig:clusters}
\end{figure}

Remarkably, one has the freedom to choose the way clusters are built in NLCEs for a given lattice geometry. For example, for a square lattice geometry, site, bond, square, and \textsf{L}-shaped building blocks~\cite{rigol_bryant_07a,rigol_bryant_07b,Abdelshafy_23,Abdelshafy_24}, and rectangular shaped clusters~\cite{Abdelshafy_24,gan_20,richter_20} have been shown to give accurate results for the thermal equilibrium properties of various spin models. For the spin lattice models studied here, we choose a rectangle expansion for the square lattice, and its corresponding parallelogram expansion for the triangular lattice, see Fig.~\ref{fig:clusters}. All clusters have the shape of a rectangle(parallelogram) with sides of length $x$ and $y$ sites for the square(triangular) lattice, and the order of the expansion $\ell$ is characterized by the clusters with the highest $x\times y$. The reasons we choose these expansion schemes are: (1) The calculation of the geometric factor $L(c)$, and consequently $M(c,\ell)$ does not involve sophisticated time-consuming combinatorial computations as simply $L(c)=2(1)$ for clusters with $x\ne y$($x=y$). (2) The number of clusters at a given order grows very slowly; The number of clusters up to a given order goes with $\sim\sqrt{\ell}$ instead of growing exponentially which is typical for expansions using building blocks, e.g. site expansion. (3) Rectangular(parallelogram) expansions made it possible to calculate the partition functions for clusters as large as $256$ sites using Bhanot's algorithm~\cite{bhanot1990numerical,creswick1995transfer}, and its extension in Ref.~\cite{sedik2024yang}. Bhanot's algorithm allows us to calculate the partition function, and hence its zeros, for large clusters as long as they are rectangular(parallelogram) and have either open or cylindrical boundary conditions. Here, we follow exactly the same algorithm that was implemented in Ref.~\cite{sedik2024yang} for the rectangular expansion, while accounting for an extra diagonal bond for the parallelogram one. 

\section{Results}
\label{Sec:Results}
In this section, we study Eq.~(\ref{eqn:rho}) for the square and the triangular lattices. The partition function in Eq.~(\ref{eqn:Z(omega)}) of a cluster $c$ is computed by exact enumeration (EE) through Bhanot's algorithm, after which the zeros $z_j(c)$ are located in the complex $z$ plane. We then present $\rho_{\ell}(z)$ by plotting the zeros $z_j(c)$ of all clusters $c$ involved in the $\ell^{th}$ order using a color map where the zeros are colored by their amplitudes $M(c,\ell)$. A color bar is included to illustrate the color mapping, and we show the results up to order $\ell=256$ in which the largest square(rhombus) cluster is $16\times 16$. The Yang-Lee zeros problem is well understood for the case of ferromagnetic (FM) interactions, so we will focus on the antiferromagnetic (AFM) interactions on the square and triangular lattice. However, we will present the FM on the square lattice to benchmark the usage of NLCE in studying Yang-Lee zeros.

\subsection{Square Lattice}
\label{SubSec:square}
\begin{figure}[t]
   \centering
   \includegraphics[width=0.99 \columnwidth]{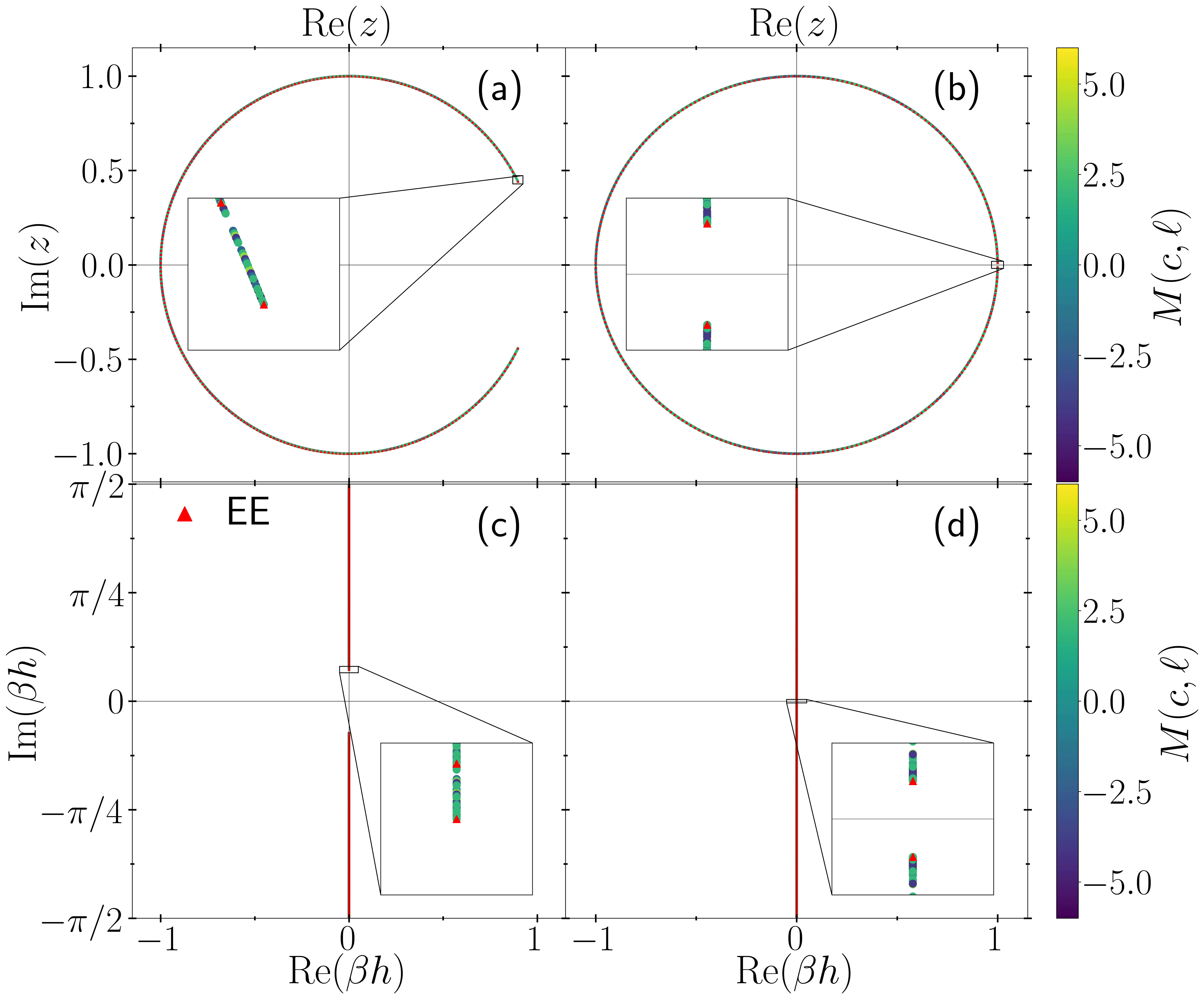}
   \caption{Yang-Lee zeros for the square lattice with ferromagnetic (FM) interactions at NLCE order $\ell=256$ for (a)$~T=2 T_c$ and (b)$~T=0.5 T_c$ in the $z$ plane, and (c)$~T=2 T_c$ and (d)$~T=0.5 T_c$ in the $h$ plane, where $T_c=\frac{2}{\ln{(1+\sqrt{2})}}$. This is contrasted with the EE zeros (red triangles) for the $16 \times 16$ cluster.}
   \label{fig:FM_sq_z_h-plane}
\end{figure}

We start by benchmarking our NLCE approach against EE calculations for the Yang-Lee zeros in the complex $z$ and $h$ planes at $\ell=256$ for the well-studied square lattice \textcolor{blue}{FM }Ising model where the Lee-Yang Circle theorem applies~\cite{lee1952statistical}. This model is known to develop long-range FM order only at zero field $h=0$ below a critical temperature $T_c$~\cite{Onsager_1944}. Above $T_c$ for $h=0$ and at any $T$ for $h\ne0$ the model is paramagnetic~\cite{lee1952statistical}. At $T>T_c=\dfrac{2}{\ln{(1+\sqrt{2})}}$ Fig.~\ref{fig:FM_sq_z_h-plane}(a), the zeros lie on an arc of the unit circle indicating that there is no phase transition. For $T\leq T_c$ Fig.~\ref{fig:FM_sq_z_h-plane}(b), the zeros populate most of the unit circle but they do not touch the positive real axis because the system is finite as shown in the inset of Fig.~\ref{fig:FM_sq_z_h-plane}(b). The zeros from NLCE fill the gaps between the zeros of the largest EE calculation ($16 \times 16$ in this case) providing a closer picture to the thermodynamic limit as shown in all the insets, which will be even more important in the AFM case. However, they do not push the Yang-Lee edge singularity closer to the real axis for any $T$ considered. 

As mentioned in Sec.~\ref{sec:free-energy-approach}, the results of Eq.~(\ref{eqn:rho}) can be expressed in the complex $\beta h$ plane as well. In Figs.~\ref{fig:FM_sq_z_h-plane}(c) and (d), we show the same zeros in Figs.~\ref{fig:FM_sq_z_h-plane}(a) and (b), respectively, in the complex $\beta h$-plane. The zeros transform as $\beta h_j(c)=\frac{1}{2}\ln\left[z_j(c)\right]$, the unit circle is mapped to the vertical line $\Re(\beta h)=0$, the disc $|z|<1(>1)$ is mapped to the half plane $\Re(\beta h)>0(<0)$, and we restrict the mapping to $ -\frac{\pi}{2}<\Im(\beta h)\leq \frac{\pi}{2}$ due to the periodicity along the imaginary axis. The same features of the zeros observed in Figs.~\ref{fig:FM_sq_z_h-plane}(a-b) are transformed to Figs.~\ref{fig:FM_sq_z_h-plane}(c-d), and it is now clearer that the phase transition happens only for $h=0$. For the remainder of this paper, we stick to presenting the results in the $h$ plane because it is easier to directly read the critical field.

\begin{figure}[t]
   \centering
   \includegraphics[width=0.98 \columnwidth]{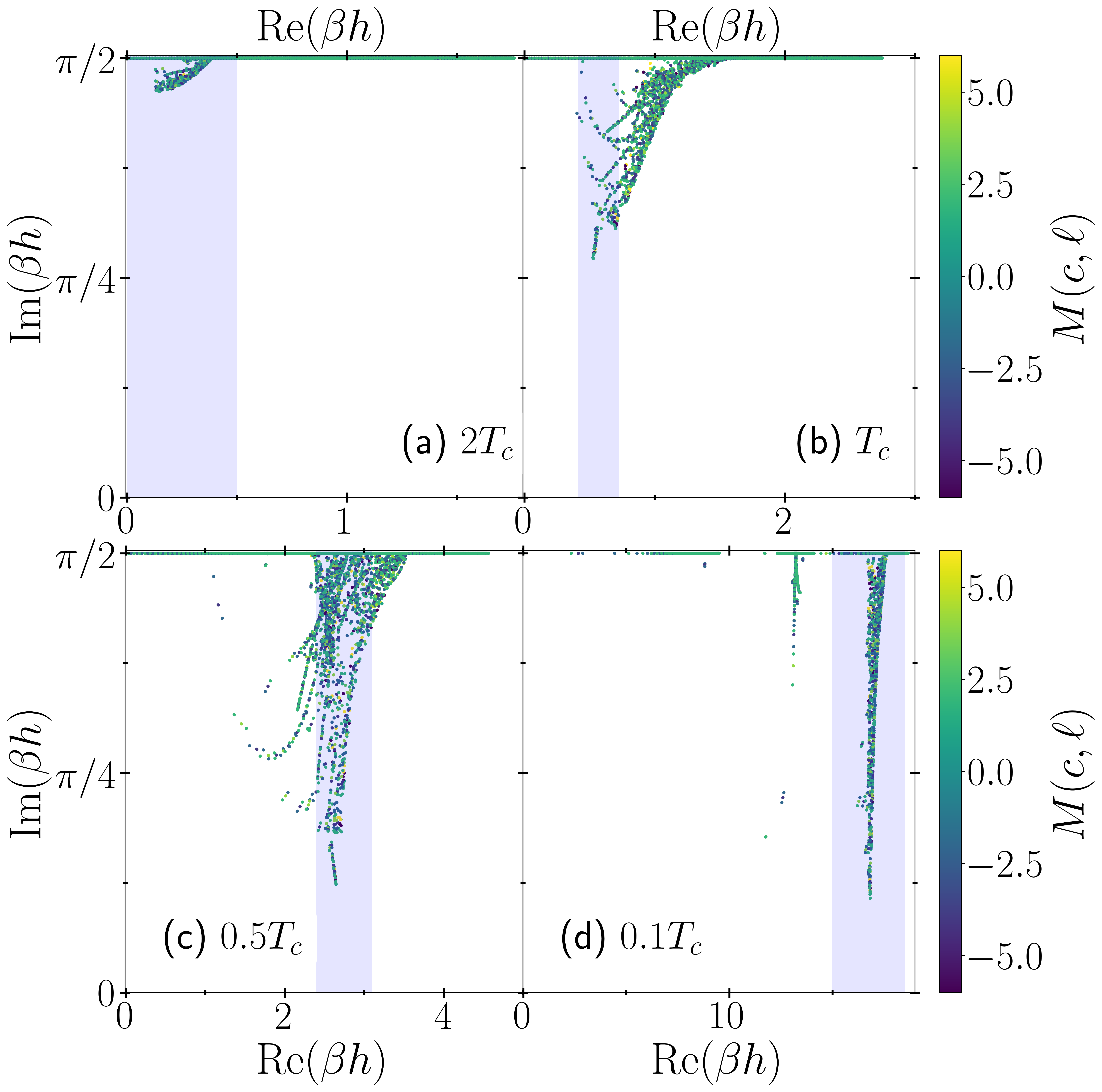}
   \caption{Yang-Lee zeros in the complex $h$ plane for the square lattice with antiferromagnetic (AFM) interactions at NLCE order $\ell=256$ for temperatures: (a)$~T=2 T_c$, (b)$~T= T_c$, (c)$~T=0.5T_c$, and (d)$~T=0.1T_c$. The region $\mathcal{D}$, dubbed a \textit{leg}, is the shaded blue vertical strip region.}
   \label{fig:AFM_sq_h-plane}
\end{figure}

The AFM square lattice Ising model at $h=0$ undergoes a phase transition at the same $T_c$, where it develops a N\'eel ordered state for $T\le T_c$~\cite{Onsager_1944}. At finite $h$, the model still develops a N\'eel ordered state for a range in $h$, but for $T<T_c$. At a sufficiently large $h$, the model is paramagnetic at any $T$ (see Fig.~\ref{fig:sq_phase-boundary}). The Yang-Lee zeros in the $h$ plane for AFM interactions are shown in Fig.~\ref{fig:AFM_sq_h-plane} for order $\ell=256$ and different $T$. Due to the symmetries of the zeros under the operations $h \leftrightarrow h^*$ and $h \leftrightarrow -h$, we restrict ourselves to the top right quadrant of the complex $h$ plane. At high temperatures (i.e. $T=2T_c$) Fig.~\ref{fig:AFM_sq_h-plane}(a), most of the zeros are on the line $\Im(\beta h)=\pi/2$ and a small fraction of the zeros are below the line but still close to it in agreement with analytical studies of the high temperature regime~\cite{heilmann1970monomers}. Upon decreasing the temperature Figs.~\ref{fig:AFM_sq_h-plane}(b-d), more zeros move away from the line $\Im(\beta h)=\pi/2$ and closer to the real axis. Moreover, most zeros move to the right, i.e. regions of larger $\Re(\beta h)$, indicating that a phase transition at lower temperatures has to take place at larger values of the external magnetic field.

In addition, the zeros start forming what we call a \textit{leg}, which is the region $\mathcal{D}$ indicated by the shaded blue vertical strip in Fig.~\ref{fig:AFM_sq_h-plane}. This region $\mathcal{D}$ covers the range $0\leq \Im(\beta h)\leq \pi/2$ and a suitable range of $\Re(\beta h)$. Specifically, we locate this range by dividing the real axis into small bins, and choose the bins with the largest weight of zeros where we assign higher weights to the zeros closer to the real axis (i.e. small imaginary part). This \textit{leg} is easier to define at lower temperatures since the variance of the real part of the zeros close to the real axis is small, but the idea of the \textit{leg} can be extended to higher temperatures by considering the region where most of the roots that are close to the real axis lie as shown for all four temperatures. We expect this region $\mathcal{D}$ to have the zeros that survive and reach the real axis in the thermodynamic limit for $T\leq T_c$. In Fig.~\ref{fig:AFM_sq_h-plane}(b-c), one can see that away from the zeros that form almost a straight line in the \textit{leg} $\mathcal{D}$, other groups of zeros are curving away from the real axis and returning back to the line $\Im(\beta h)=\pi/2$. As we show later in this paper, we argue that the latter group of zeros does not reach the real axis in the thermodynamic limit.

\begin{figure}[t]
   \centering
   \includegraphics[width=0.98 \columnwidth]{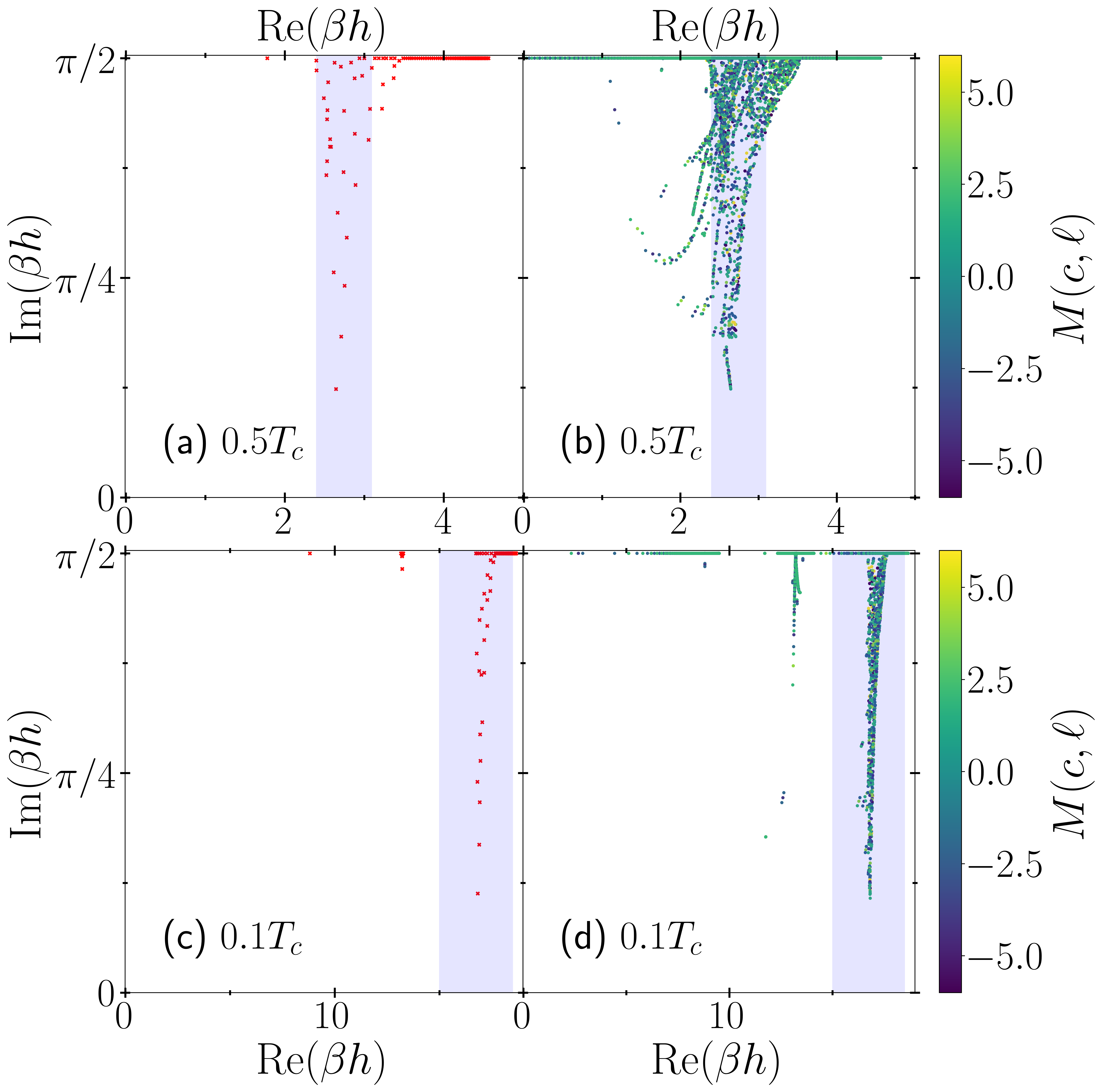}
   \caption{Yang-Lee zeros in the complex $h$ plane for the square lattice with (AFM) interactions using EE (red triangles) for the $16 \times 16$ lattice at (a)$~T=0.5 T_c$ and (c)$~T=0.1 T_c$, and NLCE at $\ell=256$ at (b)$~T=0.5 T_c$ and (d)$~T=0.1 T_c$. The shaded region is the same as in Fig.~\ref{fig:AFM_sq_h-plane}(c) and (d), respectively.}
   \label{fig:AFM_sq_h-plane_EE_vs_NLCE}
\end{figure}

Before studying the thermodynamic limit, a comparison between the distribution of the Yang-Lee zeros obtained by EE and NLCE is in order. Similar to the FM case, NLCE provides more zeros in the complex $h$ plane compared to EE with the same number of sites $\ell$ since NLCE includes zeros from other clusters with non-zero coefficients $M(c,\ell)$ in the expansion. We juxtapose the EE zeros of the $16\times 16$ cluster at $T=0.5T_c$ in Fig.~\ref{fig:AFM_sq_h-plane_EE_vs_NLCE}(a) and $T=0.1T_c$ in Fig.~\ref{fig:AFM_sq_h-plane_EE_vs_NLCE}(c) with the NLCE zeros at order $\ell=256$ for $T=0.5T_c$ in Fig.~\ref{fig:AFM_sq_h-plane_EE_vs_NLCE}(b), and $T=0.1T_c$ in Fig.~\ref{fig:AFM_sq_h-plane_EE_vs_NLCE}(d). For both temperatures, the zeros from EE and NLCE populate the same regions in the complex $h$ plane. However, the NLCE zeros show more continuous curves of zeros than EE, suggesting how the root curves look in the thermodynamic limit. In particular, at low temperatures $T\ll T_c$, the NLCE results suggest that possible root curves in the thermodynamic limit are two vertical straight lines that hit the real axis at the critical values $\pm \beta h_c$, which correspond to two concentric circles with radii $\exp(\pm2\beta h_c)$ in the complex $z$ plane.

\begin{figure}[t]
   \centering
   \includegraphics[width=0.98 \columnwidth]{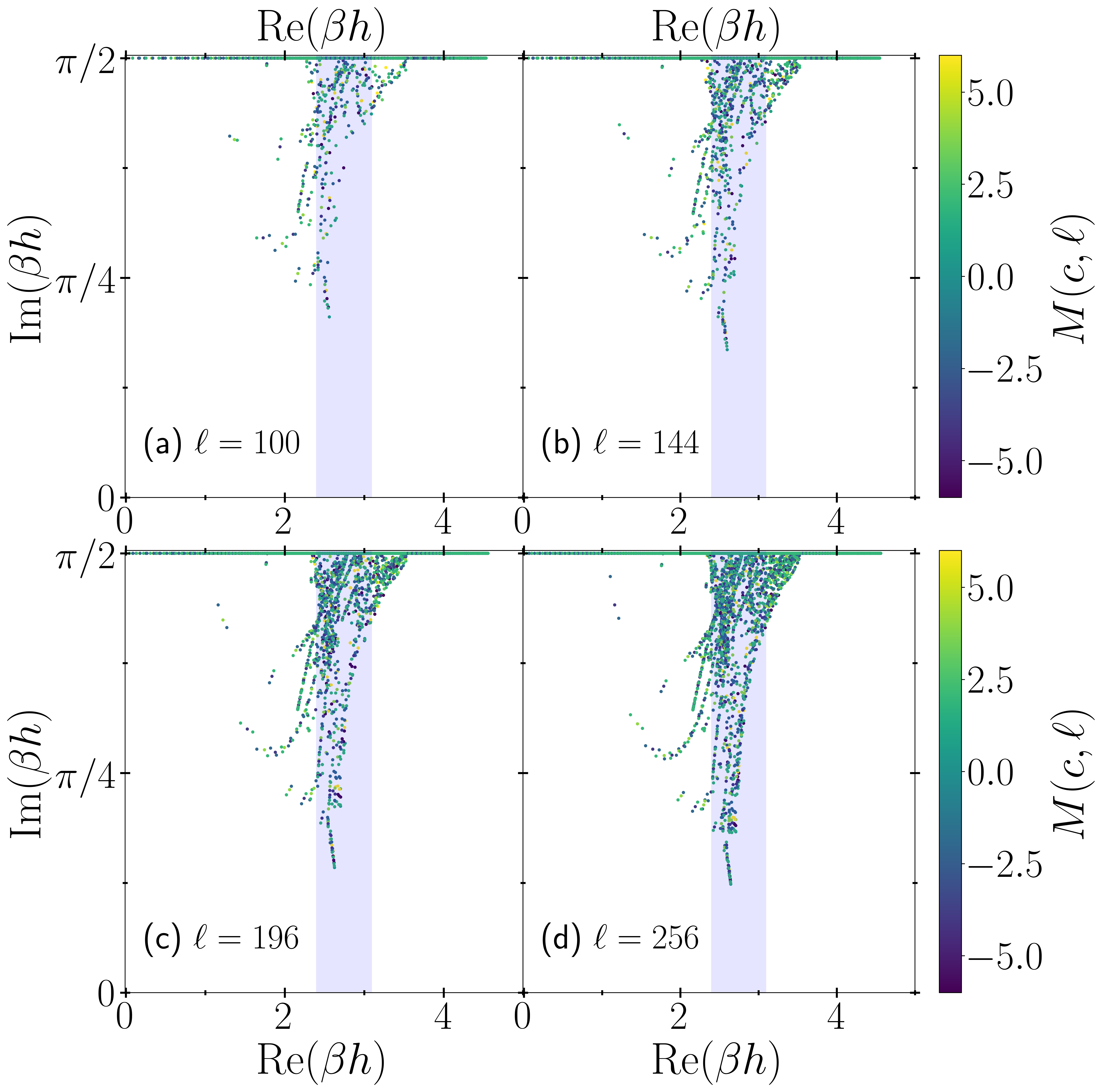}
   \caption{Yang-Lee zeros in the complex $h$ plane for the square lattice with antiferromagnetic (AFM) interactions at $T=0.5T_c$ using NLCE at different orders: (a)$~\ell=100$, (b)$~\ell=144$, (c)$~\ell=196$, and (d)$~\ell=256$. The shaded region $\mathcal{D}$ (\textit{leg}) is determined by the zeros in the largest computed order $\ell=256$. The shaded region is the same as in Fig.~\ref{fig:AFM_sq_h-plane}(c).}
   \label{fig:AFM_sq_h-plane-fixed-T}
\end{figure}

To assess the behavior of the Yang-Lee zeros as the thermodynamic limit is approached, we study the zeros at $T=0.5T_c$ and different NLCE order $\ell$ in Fig.~\ref{fig:AFM_sq_h-plane-fixed-T}. The shaded region $\mathcal{D}$ (\textit{leg}) is defined using the zeros in the largest computed order $\ell=256$, and applied for all lower orders. As the NLCE order increases, the zeros in the \textit{leg} move closer to the real axis, and align more in a straight line. Nevertheless, the other group of zeros does not approach the real axis with increasing the NLCE order $\ell$. These observations suggest that in the thermodynamic limit, only those in the \textit{leg} $\mathcal{D}$ would survive and touch the real axis. A similar behavior is observed in all the other temperatures tried.

\begin{figure}[b]
   \centering \includegraphics[width=0.98\columnwidth]{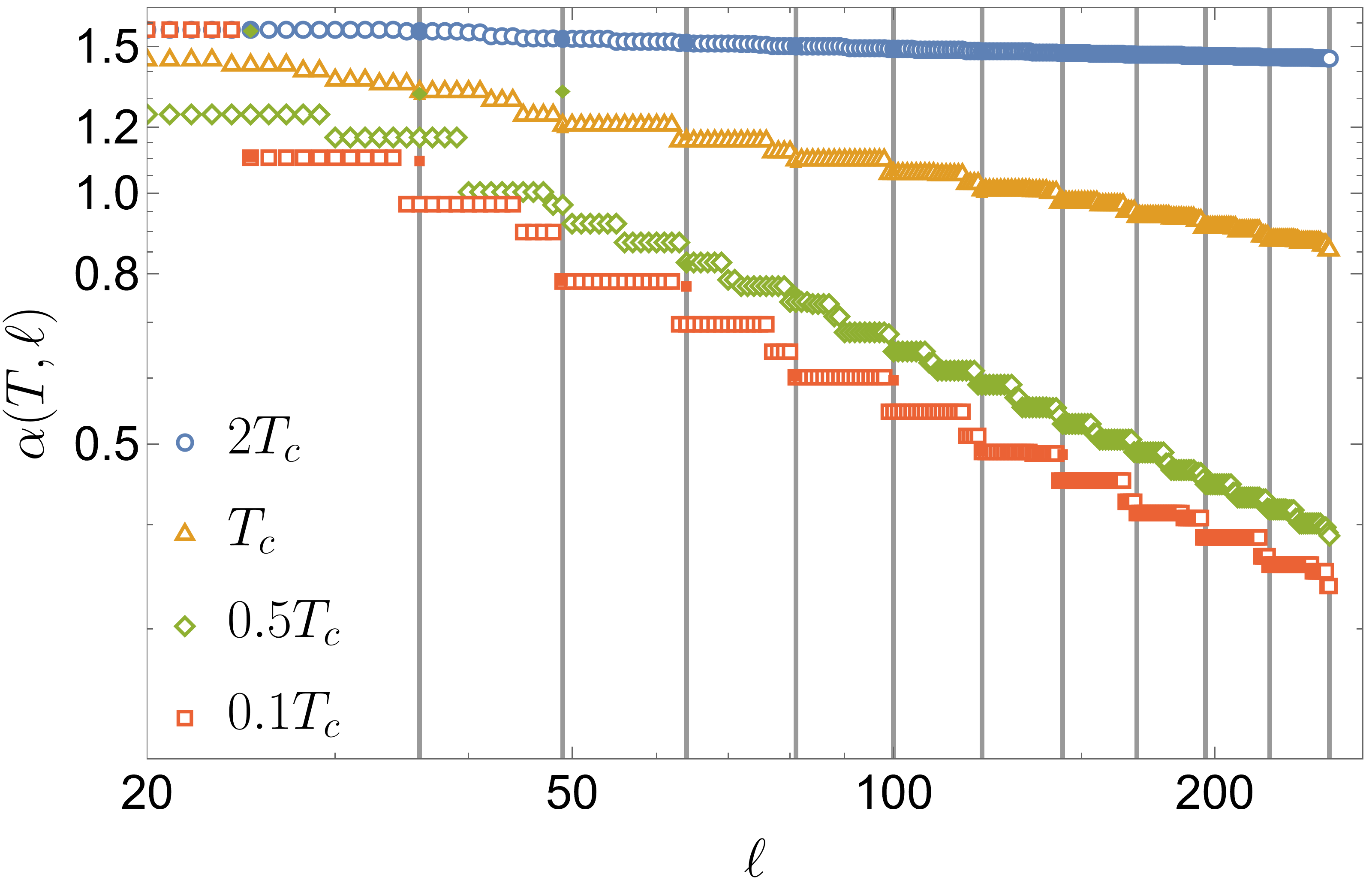}
   \caption{Log-log scale of $\alpha(T,\ell)$ vs $\ell$ for NLCE (open markers) and EE (filled markers) at temperatures $T=2T_c$, $T=T_c$, $T=0.5T_c$, and $T=0.1T_c$ for the square lattice with (AFM) interactions. EE has values only at $\ell$s that admit perfect square clusters, which are indicated by the vertical grid lines. The polynomial convergence of the NLCE and EE for $T\leq T_c$ is indicated by the linear trend.}
   \label{fig:sq_conv}
\end{figure}

After establishing why the zeros in the \textit{leg} are the ones of interest from the physics point of view, we quantify the convergence of the NLCE and EE results to the corresponding thermodynamic limit. We base this on the imaginary part of the closest zero to the real axis in the \textit{leg} $\mathcal{D}$, which depends on the temperature $T$ and the order $\ell$ and defined by
\begin{align}
    \alpha(T,\ell) \equiv \min_{\beta h \in \mathcal{D}}\left[\Im(\beta h)\right].
    \label{eqn:alpha-def}
\end{align}
The smaller $\alpha(T,\ell)$ (i.e. the closer the zeros are to the real axis), the closer our zeros to the thermodynamic picture we seek, and by this definition of convergence, NLCE and EE results can be compared. This comparison is shown in Fig.~\ref{fig:sq_conv} by plotting $\alpha(T,\ell)$ as a function of the NLCE order $\ell$ for 4 different temperatures. The straight line dependence for $T\leq T_c$ shows a polynomial convergence to the thermodynamic limit of both the NLCE and EE results as a function of the NLCE order. This is to be contrasted with the exponential convergence observed for NLCEs calculations of thermodynamic observables~\cite{iyer_15,Abdelshafy_23}. Apart from some deviations for EE calculations for small square clusters ($\ell \le 100$) with even sides --since they cannot host a perfect antiferromagnetic-ordered state--, the convergence of both EE and NLCE show similar trend and proximity to the thermodynamic limit at a given temperature. We can also see that $T=2T_c$ saturates with increasing $\ell$ implying no phase transition at this temperature. We thus conclude that NLCE and EE are quite similar when studying how fast the closest zero touches the real axis. 

Knowing how the distribution of zeros approaches the thermodynamic limit, one can try to extrapolate the value of $\Re(\beta h)$ at which this straight line would touch the real axis. This is done by considering the closest $m$ zeros to the real axis in the region $\mathcal{D}$ (\textit{leg}) for a given $T$, finding their best-fit line, then extrapolating the value $h_c(T)$. The inset in Fig.\ref{fig:sq_phase-boundary}(a) shows an example of the extrapolation process where the best-fit line is drawn using the closest NLCE $m=11$ zeros to the real axis for $T=0.5T_c$. By doing this for different values of $T$, one could plot $h_c(T)$ using the NLCE zeros and the EE zeros in comparison with the phase boundary $h_{c;\text{WK}}$ obtained by Wang and Kim \cite{wang1997critical}, which agrees with other results in literature \cite{muller1977interface,monroe2001systematic,shore2015charge}.  Such a comparison for $m=11$ is shown in Fig.~\ref{fig:sq_phase-boundary}(a), where we see that the extrapolated phase boundary using NLCE is closer to the WK one compared to the EE. This comparison is quantified as follows. We consider the mean relative error (across all temperatures $T<T_c$) between the extrapolated $h_c(T)$ and the literature value value defined by 
 \begin{align}
     \Delta= \underset{T<T_c}{\mathrm{mean}}\left[\frac{\left|h_c(T)-h_{c;\text{WK}}(T)\right|}{h_{c;\text{WK}}(T)}\right].
     \label{eqn:error-def}
 \end{align}

The mean relative error $\Delta$ for NLCE extrapolation is then compared to the mean relative error for EE. In Fig.~\ref{fig:sq_phase-boundary}(b), we show the dependence of the mean relative error $\Delta$ on the number of points $m$ included in the best-fit line for NLCE and EE. The NLCE extrapolation has a smaller mean relative error for all $m$ we tried compared to the EE extrapolation because NLCE provides more zeros as seen in Fig.~\ref{fig:AFM_sq_h-plane_EE_vs_NLCE}. It is important to note that the accuracy of the NLCE's phase boundary is robust for a long range of choices of $m$ since $\Delta$ remains very small ($\lesssim 5\%$) for $3\le m \le 35$. Both the EE and NLCE errors go up after a sufficiently large $m$ because zeros close to the $\Im(\beta h) = \pi/2$ line start to be included in the linear fit. However, since NLCE has more zeros in the leg, their fit is more accurate and robust over a larger range of $m$.

\begin{figure}[t]
   \centering
   \includegraphics[width=0.98\columnwidth]{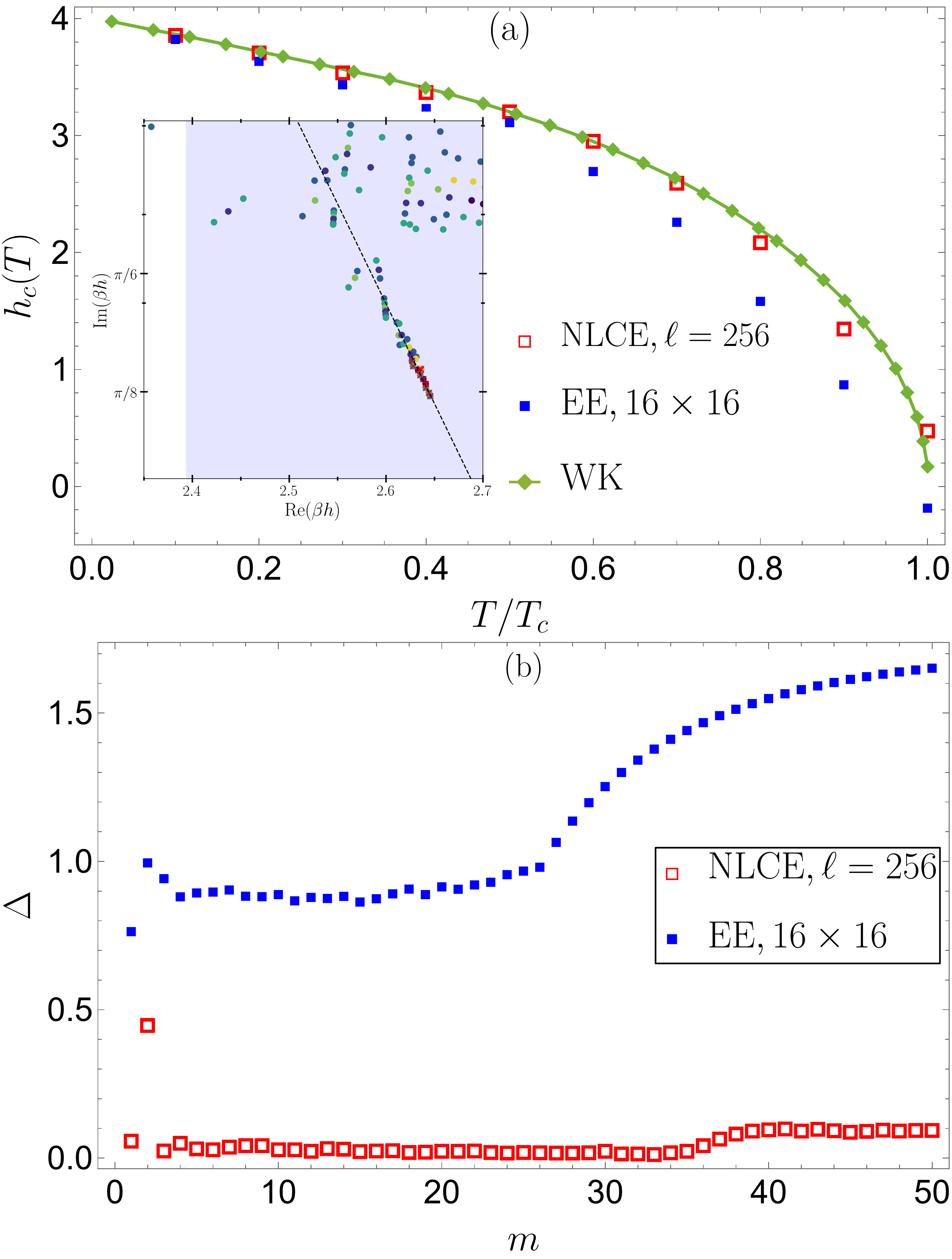}
   \caption{Panel (a) shows the extrapolated phase boundary for the square lattice with (AFM) interactions through a best-fit line of the closest 11 NLCE (red open squares) zeros in the \textit{leg} contrasted with the closest 11 EE (blue filled squares) zeros to the real axis. i.e. $m=11$ is used for this plot. The inset shows an example of the best-fit line (dashed line) and the closest 11 NLCE zeros to the real axis (red crosses) for $T=0.5 T_c$, which is a zoomed-in version of Figs.~\ref{fig:AFM_sq_h-plane}(c) and \ref{fig:AFM_sq_h-plane-fixed-T}(d). The phase boundary obtained by Wang and Kim \cite{wang1997critical} is shown in green diamonds. The mean relative error across all temperatures $\Delta$, defined in Eq.~(\ref{eqn:error-def}), is plotted against different values of $m$ in (b) , where the error from NLCE extrapolation is smaller than that from EE for all values of $m$.}
   \label{fig:sq_phase-boundary}
\end{figure}
It is also interesting to mention that for the case $m=1$, where the extrapolation is merely taking the real part of the closest zero to the real axis, NLCE still outperforms EE. This occurs because the closest zero in the region $\mathcal{D}$ (\textit{leg}) does not come from the largest perfect square cluster, it rather comes from a different rectangular cluster. For example, at $T=0.1T_c$ and order $\ell=256$, the zero closest to the real axis comes from the rectangular cluster $15\times 17 = 255$ not from the square $16 \times 16=256$ cluster.

\subsection{Triangular}\label{SubSec:triangular}
\begin{figure}[t]
    \centering
    \includegraphics[width=0.98 \columnwidth]{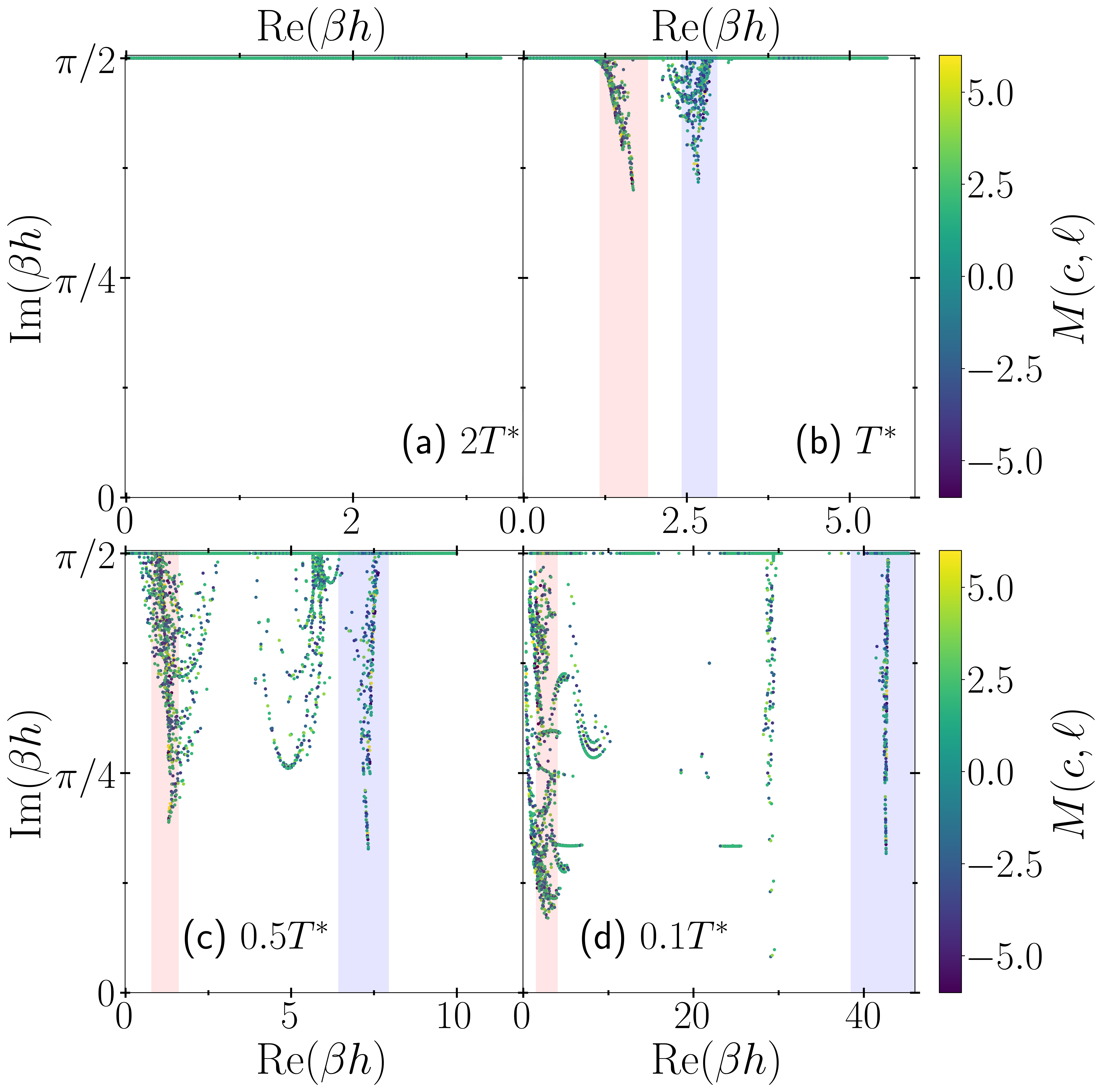}
    \caption{Yang-Lee zeros in the complex $h$ plane for the triangular lattice with antiferromagnetic (AFM) interactions at NLCE order $\ell=256$ for temperatures: (a)$~T=2 T^*$, (b)$~T= T^*$, (c)$~T=0.5T^*$, and (d)$~T=0.1T^*$, where $T^*\approx1.361$. The region $\mathcal{D_{\text{left}}}$($\mathcal{D_{\text{right}}}$), is the shaded red(blue) vertical strip region.}
    \label{fig:AFM_tr_h-plane}
\end{figure}

\begin{figure}[t]
   \centering
   \includegraphics[width=0.98\columnwidth]{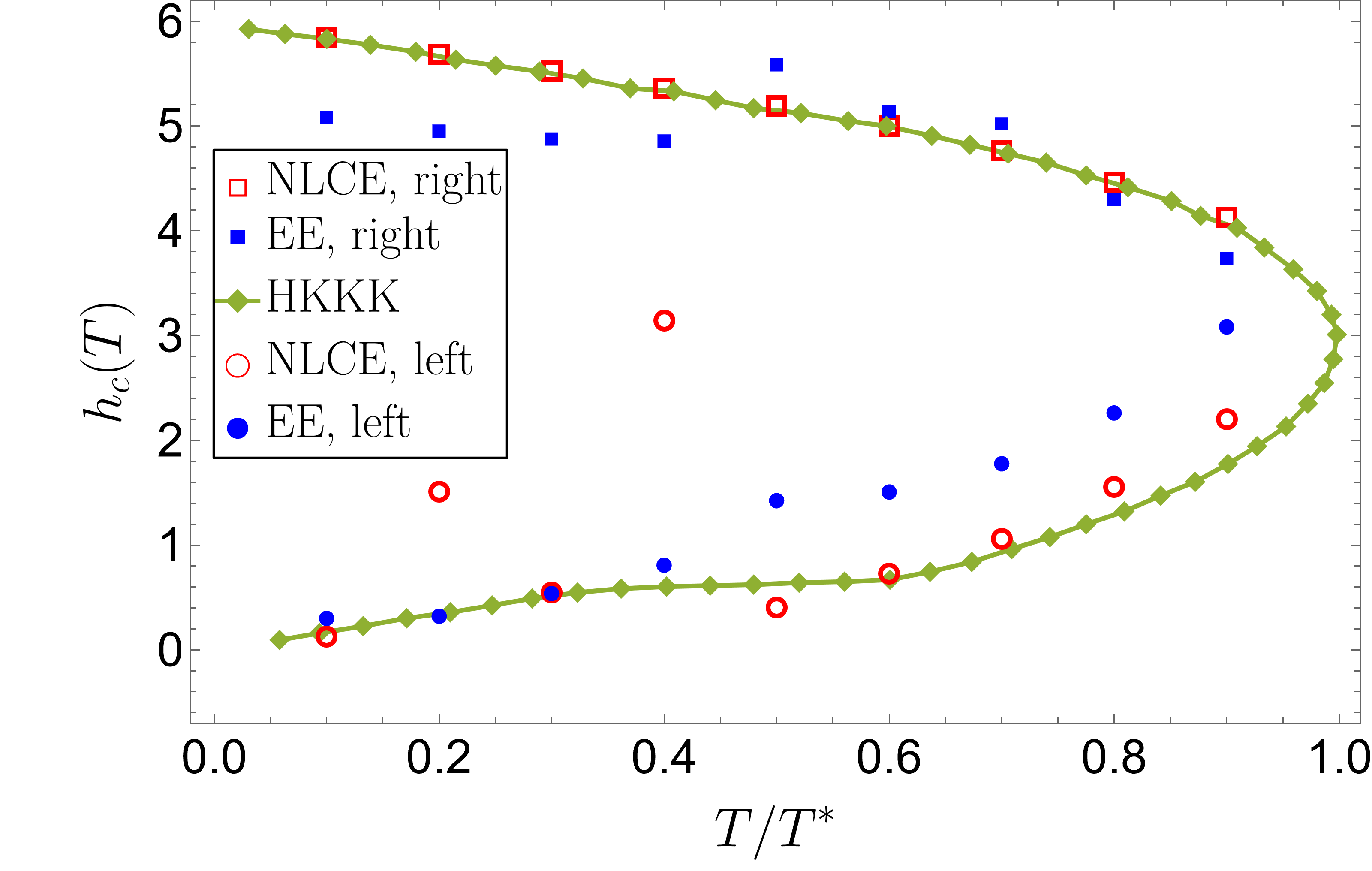}
   \caption{The extrapolated phase boundary for the triangular lattice with (AFM) interactions as red open circles(squares) through a best-fit line of the closest 16(8) NLCE zeros to the real axis in the left(right) leg $\mathcal{D}_{\text{left}}(\mathcal{D}_{\text{right}})$ and the extrapolated phase boundary blue filled circles(squares) with closest 16(8) EE zeros to the real axis. i.e. $m_{\text{left}}=16$ and $m_{\text{right}}=8$ is used for this plot, while the phase boundary obtained in Ref.~\cite{hwang2008thermodynamic} is shown in green diamonds.}
   \label{fig:tr_phase-boundary}
\end{figure}

Next, we turn to the AFM Ising model on a triangular lattice, which is a frustrated model with a highly degenerate ground state and no long-range order at zero field $h=0$~\cite{Wannier_1950,HOUTAPPEL1950425}. The introduction of a longitudinal field, i.e. when $h\ne0$,  lifts that degeneracy but still favors an antiferromagnetic alignment of the spins~\cite{schick1977phase}. The ground state becomes only three-fold degenerate where two of the three spins in each triangle unit are aligned with the external field. In this region, the models goes through a phase transition as the temperature increases from this antiferromagnetic phase to a paramagnetic phase~\cite{schick1977phase}. At sufficiently high fields, this model becomes paramagnetic at all temperatures, namely, all spins align with the external field. We present the results of $\rho_{\ell}(z)$ in Eq.~(\ref{eqn:rho}) in the same way as we did for the square lattice. Again, we restrict ourselves to the top right quadrant of the complex $h$ plane because of the symmetries $h \leftrightarrow h^*$ and $h \leftrightarrow -h$. Fig.~\ref{fig:AFM_tr_h-plane} shows Yang-Lee zeros in the $h$ plane for AFM interactions at NLCE order $\ell=256$ and different $T$. The temperature $T^*\approx 1.1361$ is the highest temperature at which there is a phase transition in the system~\cite{noh1992phase,hwang2008thermodynamic,hwang2010yang}, see Fig.~\ref{fig:tr_phase-boundary}. At high temperatures (i.e. $T=2T^*$) Fig.~\ref{fig:AFM_tr_h-plane}(a), all the zeros lie on the line $\Im(\beta h)=\pi/2$ as predicted. As the temperature decreases Figs.~\ref{fig:AFM_tr_h-plane}(b-d), more zeros move closer to the real axis and away from the line $\Im(\beta h)=\pi/2$. Opposite to the square lattice, where most zeros move to the right, the zeros expand to cover a wider region of the real axis.

The zeros departing from the $\Im(\beta h)=\pi/2$ split into sets, one of which moves to the right and another moves to the left. We therefore can extend our definition of \textit{leg} introduced for the AFM square lattice to the case of triangular lattice. We define the left(right) \textit{leg} by the region $\mathcal{D}_{\text{left}}$($\mathcal{D}_{\text{right}}$), which is the region indicated by the shaded red (blue) vertical strip in Fig.~\ref{fig:AFM_tr_h-plane}. $\mathcal{D}_\text{left}$($\mathcal{D}_{\text{right}}$) covers $0\leq \Im(\beta h)\leq \pi/2$ and a range of $\Re(\beta h)$ that is chosen in the same manner as explained in the case of a square lattice. In agreement with the square lattice, the \textit{legs} are easier to define at lower temperatures. We, again, expect these \textit{legs} to have the zeros that survive and reach the real axis in the thermodynamic limit for $T\ll T^*$, see Fig.~\ref{fig:AFM_tr_h-plane}(d). As $T$ increases towards $T^*$, although the zeros in the \textit{legs} are still distinct, the two sets of zeros come closer to each other and the description of having two distinctive legs becomes less accurate and can break down. This is particularly clear for the case $T=T^*$, Fig.~\ref{fig:AFM_tr_h-plane}(b) where the zeros are expected to pinch the real axis at a single point. Hence, we believe the apparent splitting into two legs may be due to finite-size effects. The zeros that are outside the \textit{legs} in Fig.~\ref{fig:AFM_tr_h-plane}(b-c), are expected to not reach the real axis in the thermodynamic limit similar to the ones we saw in the square lattice, as we show later. 

\begin{figure}[t]
   \centering
   \includegraphics[width=0.98 \columnwidth]{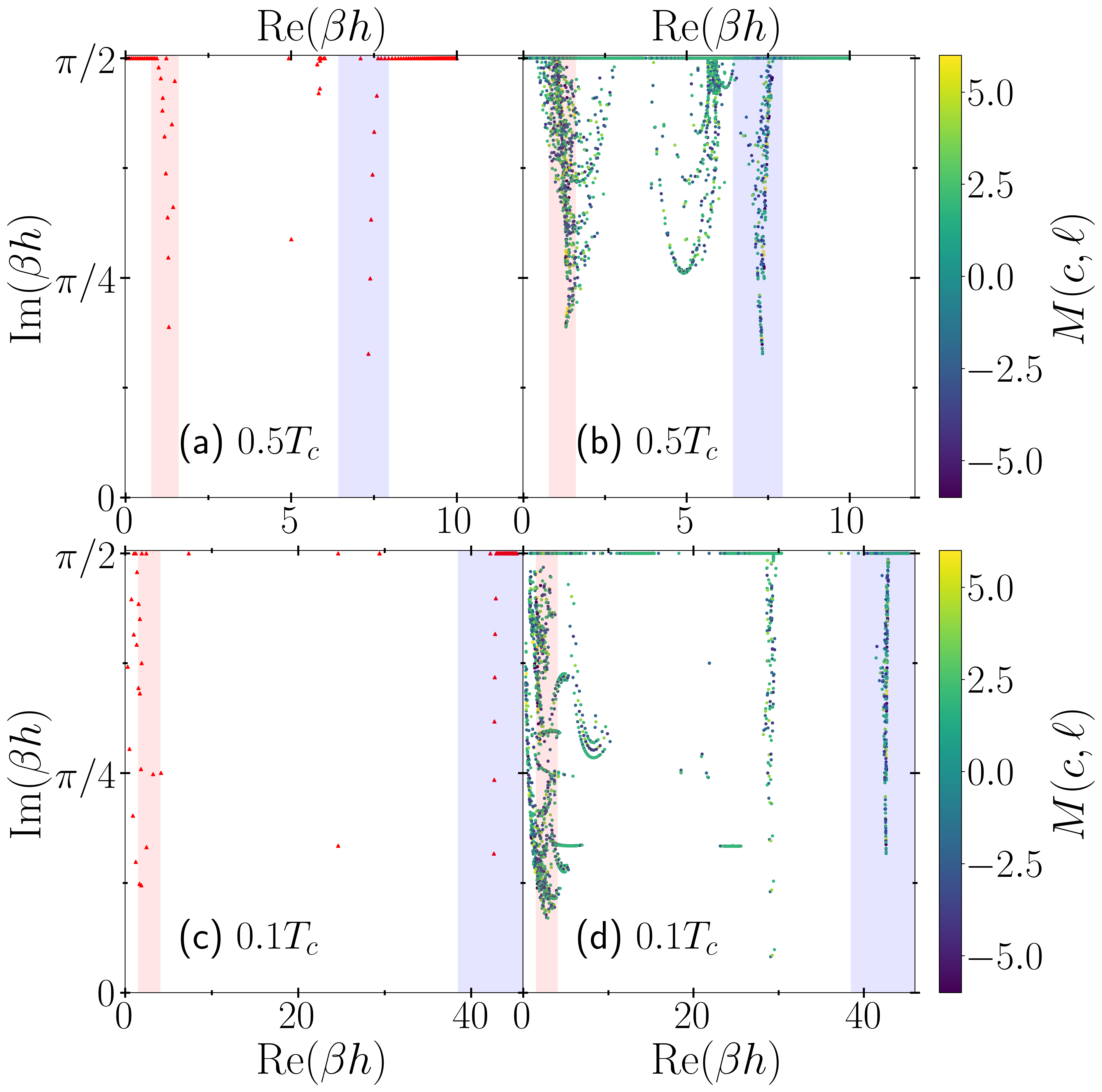}
   \caption{Yang-Lee zeros in the complex $h$ plane for the triangular lattice with (AFM) interactions using EE (red triangles) for the $16 \times 16$ lattice at (a)$~T=0.5 T^*$ and (c)$~T=0.1 T^*$, and NLCE at $\ell=256$ at (b)$~T=0.5 T^*$ and (d)$~T=0.1 T^*$. The shaded regions are the same as in Fig.~\ref{fig:AFM_tr_h-plane}(c) and (d), respectively.}
   \label{fig:AFM_tr_h-plane_EE_vs_NLCE}
\end{figure}

We first draw a comparison between the Yang-Lee zeros distributions obtained from EE and NLCE. We compare the EE zeros of the $16\times 16$ cluster at $T=0.5T^*$ in Fig.~\ref{fig:AFM_tr_h-plane_EE_vs_NLCE}(a) and $T=0.1T^*$ in Fig.~\ref{fig:AFM_tr_h-plane_EE_vs_NLCE}(c) to the NLCE zeros at order $\ell=256$ for $T=0.5T^*$ in Fig.~\ref{fig:AFM_tr_h-plane_EE_vs_NLCE}(b), and $T=0.1T^*$ in Fig.~\ref{fig:AFM_tr_h-plane_EE_vs_NLCE}(d). NLCE provides many more zeros in the complex $h$ plane compared to EE with the same number of sites $\ell$ by the same reasoning we showed in the square lattice. The zeros from EE and NLCE lie in the same regions in the complex $h$ plane for both temperatures, but the NLCE zeros present a more continuous picture of the zeros compared to EE, which hints at how the thermodynamic root curves could be. For $T\ll T^*$, the zeros in the right \textit{leg} $\mathcal{D}_{\text{right}}$ form a clear vertical line suggesting that the thermodynamic root curves for this phase transition are two vertical lines that pinche the real axis at critical value $\pm \beta h_c$, which correspond to two circles with radii $\exp(\pm \beta h_c)$ in the complex $z$ plane. On the other hand, the zeros in the left \textit{leg} $\mathcal{D}_{\text{left}}$ do not form as clear of a straight line, which could be due to finite size effects since the correlation length increases with decreasing temperatures at low values of external magnetic field $\Re(h)$ \cite{stephenson1964ising,jacobsen1997monte,kim2015ising}.

\begin{figure}[t]
    \centering
    \includegraphics[width=0.98 \columnwidth]{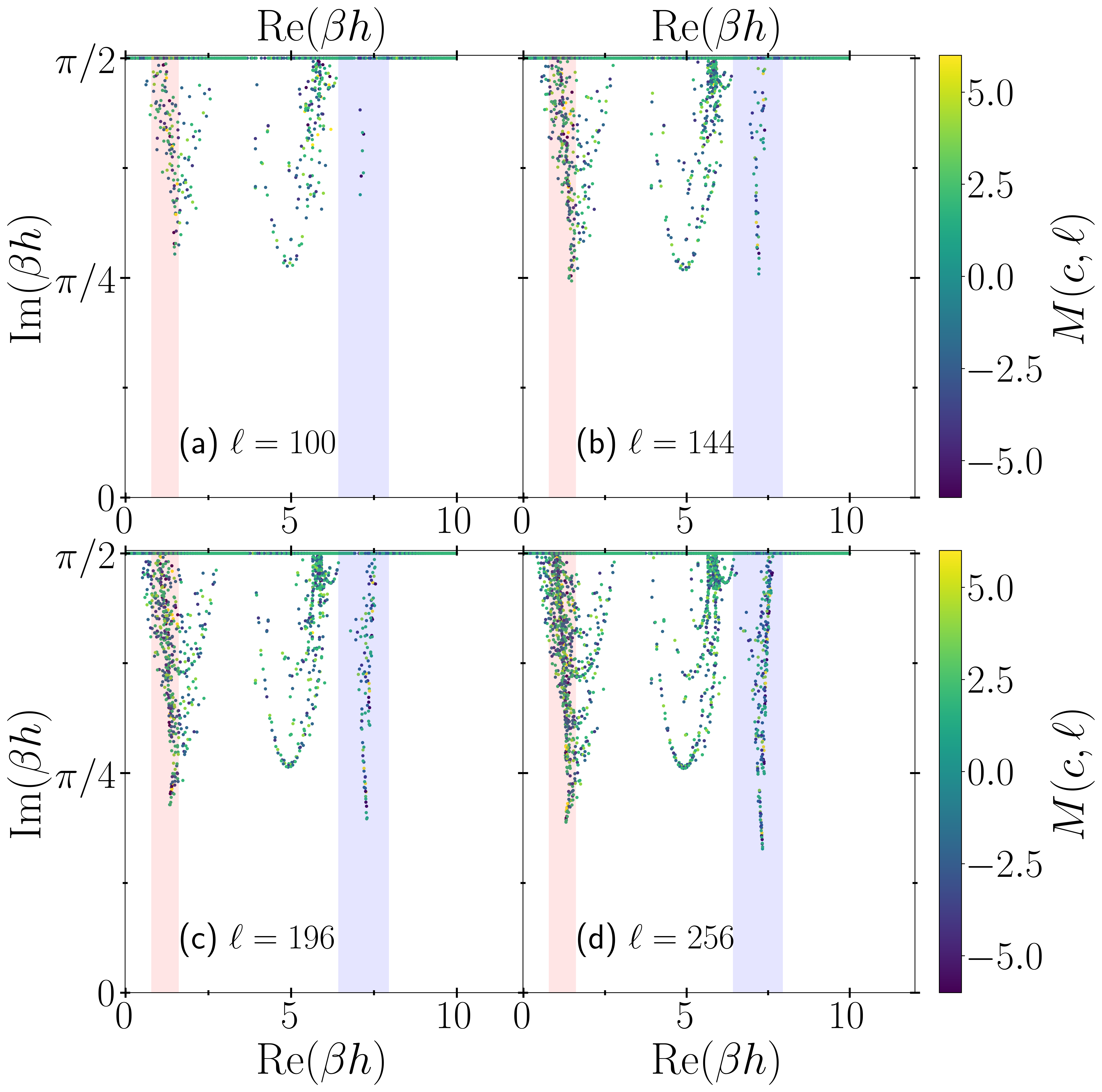}
    \caption{Yang-Lee zeros in the complex $h$ plane for the square lattice with antiferromagnetic (AFM) interactions at $T=0.5T^*$ using NLCE at different orders: (a)$~\ell=100$, (b)$~\ell=144$, (c)$~\ell=196$, and (d)$~\ell=256$. The shaded regions are the same as in Fig.~\ref{fig:AFM_tr_h-plane}(c).}
    \label{fig:AFM_tr_h-plane-fixed-T}
\end{figure}

We now study the zeros at $T=0.5T^*$ and different NLCE order $\ell$ in Fig.~\ref{fig:AFM_tr_h-plane-fixed-T} to gain insights about the thermodynamic limit picture of the root curves. The \textit{legs} $\mathcal{D}_\text{left}$ ($\mathcal{D}_\text{right}$) are defined using the zeros in the largest computed order $\ell=256$, and applied for all lower orders. Upon increasing the NLCE order, the zeros in each \textit{leg} move closer to the real axis, while the zeros outside of the legs retain a fixed minimum imaginary part with increasing $\ell$, similar to our observations in the square lattice. We therefore expect that only the zeros in the \textit{legs} survive and pinch the real axis in the thermodynamic limit. For all the temperatures we tried, the zeros exhibit the same behavior.

\begin{figure}[t]
   \centering
   \includegraphics[width=0.98\columnwidth]{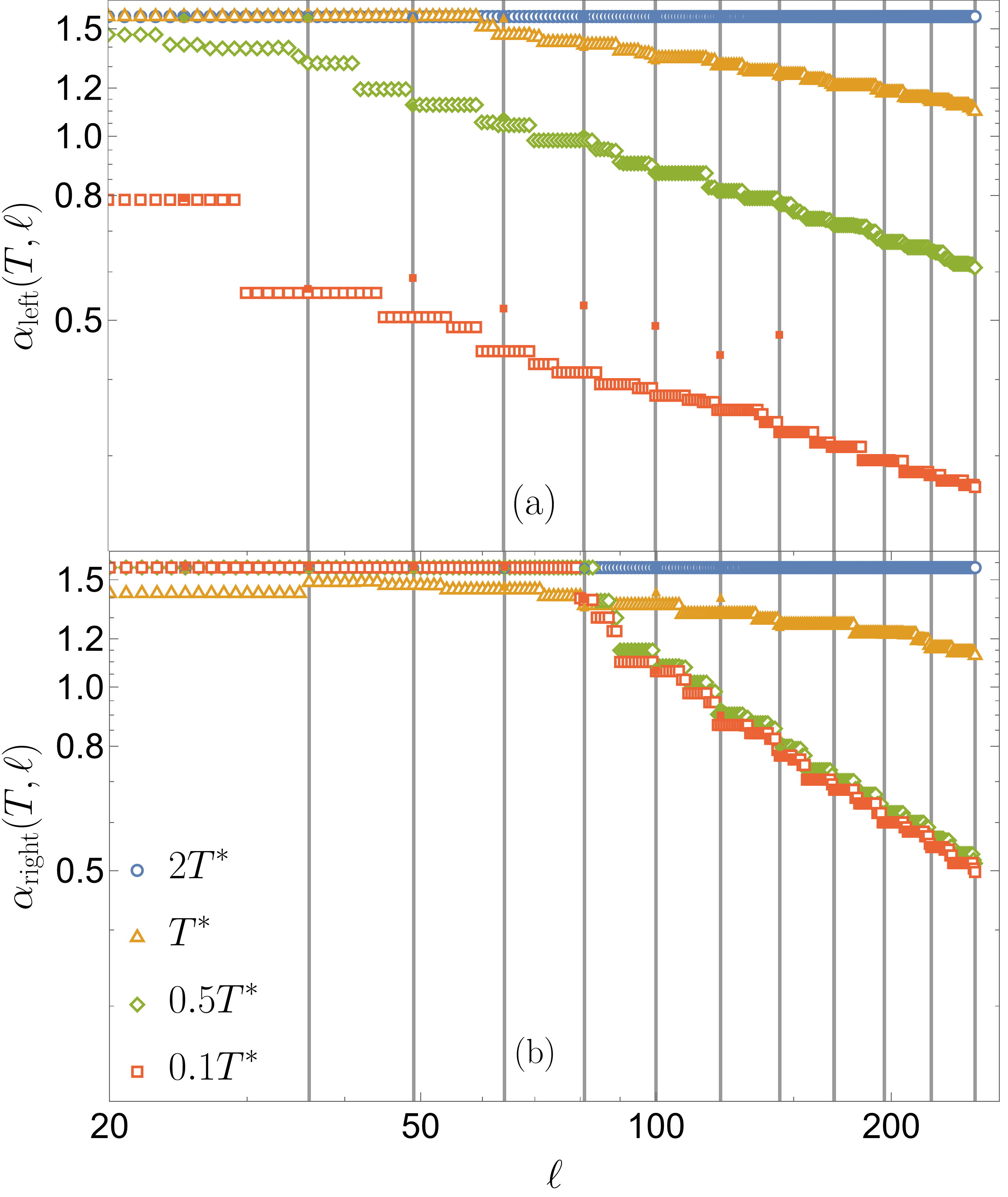}
   \caption{Log-log scale of (a) $\alpha_{\text{left}}(T,\ell)$ (b) $\alpha_{\text{right}}(T,\ell)$ vs $\ell$ for NLCE (open markers) and EE (filled markers) at temperatures $T=2T^*$, $T=T^*$, $T=0.5T^*$, and $T=0.1T^*$ for the triangular lattice with (AFM) interactions. EE has values only at $\ell$s that admit perfect rhombus clusters, which are indicated by the vertical grid lines. The polynomial convergence of the NLCE and EE for $T\leq T^*$ is indicated by the linear trend.}
   \label{fig:tr_conv}
\end{figure}

We study the convergence of the NLCE and EE results to the thermodynamic limit in the same manner as the square lattice. We consider the imaginary part of the closest zero in each \textit{leg} $\mathcal{D}_{\text{left}}$($\mathcal{D}_{\text{left}}$), which depends on the temperature $T$ and the order $\ell$ and defined, similar to Eq.~(\ref{eqn:alpha-def}), by
\begin{align}
    \alpha_{(\cdot)}(T,\ell) \equiv \min_{\beta h \in \mathcal{D}_{(\cdot)}}\left[\Im(\beta h)\right],
    \label{eqn:alpha-tr}
\end{align}
where $(\cdot)$ is either left or right. The smaller $\alpha(T,\ell)$ (i.e. the closer the zeros are to the real axis), the closer our zeros to the thermodynamic picture we seek, and by this definition of convergence, NLCE and EE results can be compared. This comparison is shown in Fig.~\ref{fig:tr_conv} by plotting (a) $\alpha_{\text{left}}(T,\ell)$  and (b) $\alpha_{\text{right}}(T,\ell)$ as functions of the NLCE order $\ell$. The figure shows the $\log$-$\log$ scale of these relations and different temperatures $T$ are plotted with different legends for NLCE (Open markers) and EE (filled markers) shown at vertical grid-lines corresponding to when $\ell$ is a perfect square. The polynomial convergence to the thermodynamic limit of both the NLCE and EE results is deduced from the linear dependence for $T< T^*$ as a function of the NLCE order, which is the same dependence found in the square case. We also observe that the convergence of the right \textit{leg} for $T\leq T^*$ is similar for different temperatures, which was the case in the square lattice. However, it appears that the convergence of the left \textit{leg} worsens as $T$ increases towards $T^*$ indicating that one may need larger system size to better capture any phase transition happening at the left \textit{leg}. The fact that all zeros are on the line $\Im(\beta h)=\pi/2$ for $T>T^*$ is reflected in the constant dependence with increasing $\ell$ indicating no phase transition at these temperatures. For $T=T^*$, we can see that $\alpha_{\text{left}}\neq\alpha_{\text{right}}$ even though they are expected to pinch the real axis at the same point. We suspect that this is due to finite size effects and we, therefore, refrain from using $T^*$ for further analysis.

Since the zeros in the right \textit{leg} $\mathcal{D}_{\text{right}}$ form a straight line, one can try to extrapolate the value of $\Re(\beta h)$ at which this straight line would touch the real axis. We also apply the same extrapolation to the zeros in the left \textit{leg} $\mathcal{D}_{\text{left}}$. Similar to the square lattice, we consider the closest $m$ zeros to the real axis in a given \textit{leg} for a given $T$, then extrapolating the value $h_c(T)$ at which their best-fit line pinches the real axis. Since the extrapolation is done for each \textit{leg} separately, one can choose different number of zeros $m_{\text{left}}$($m_{\text{right}}$) in a given leg. For example, Fig.~\ref{fig:tr_phase-boundary} shows the extrapolated phase boundary $h_c(T)$ using $m_{\text{left}}=16$ and $m_{\text{right}}=8$. On the other hand, the EE zeros are split into two parts around the midpoint given by $\left\{\max_{j}\left[\Re(\beta h_j)\right]+\min_{j}\left[\Re(\beta h_j)\right]\right\}/2$. The extrapolation from the EE zeros on the left(right) of the midpoint is compared to the extrapolation from NLCE zeros in region $\mathcal{D}_{\text{left}}$($\mathcal{D}_{\text{right}}$) and to the phase boundary $h_{c;\text{HKKK}}$ obtained using the Wang-Landau Monte Carlo algorithm by Hwang-Kim-Kang-Kim \cite{hwang2008thermodynamic}, and similar results in literature \cite{schick1977phase,noh1992phase,monroe1998frustrated}, in Fig.~\ref{fig:tr_phase-boundary}. The extrapolated $h_c(T)$ from NLCE zeros in the right \textit{leg} $\mathcal{D}_{\text{right}}$ are quite close to the HKKK values and are much better than the corresponding EE extrapolations. However, the extrapolations from the NLCE zeros in the left \textit{leg} $\mathcal{D}_{\text{left}}$ worsens with decreasing $T$ and the corresponding $h_{c;HKKK}$. 

\begin{figure}[t]
   \centering
   \includegraphics[width=0.98\columnwidth]{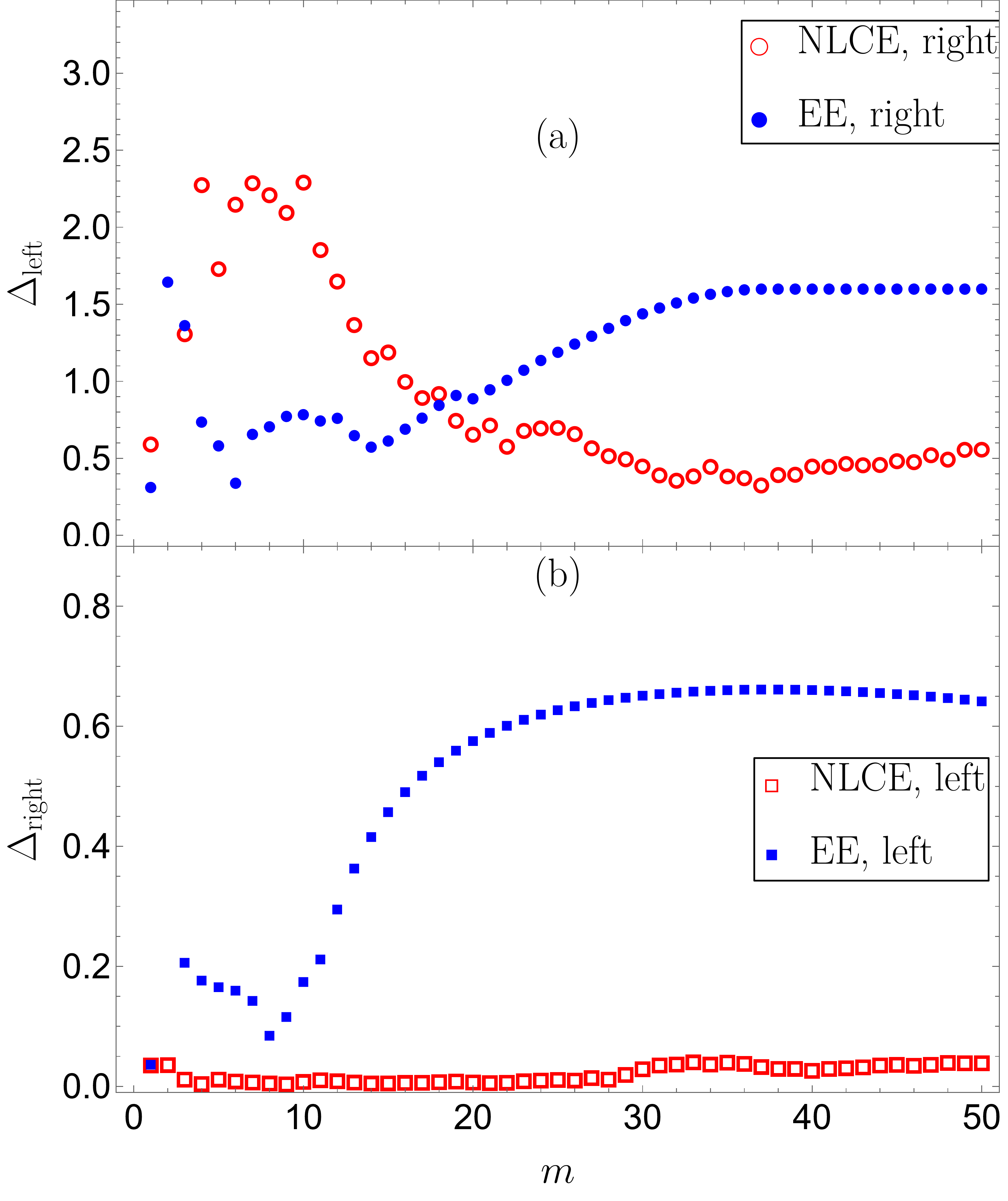}
   \caption{The mean relative error across all temperatures $T<T^*$ for the triangular lattice with (AFM) interactions for (a) left leg $\Delta_{\text{left}}$ and (b) right leg $\Delta_{\text{right}}$, defined in Eq.~(\ref{eqn:error-tr}), is plotted against different values of $m$.}
   \label{fig:tr_phase-boundary-err}
\end{figure}

We again quantify the comparison by considering the mean relative error (across all temperatures $T<T^*$) between the extrapolated $h_c(T)$ and the HKKK value defined by 
\begin{align}
    \Delta_{(\cdot)}= \underset{T<T^*}{\mathrm{mean}}\left[\frac{\left|h_c(T)-h_{c;\text{HKKK}}(T)\right|}{h_{c;\text{HKKK}}(T)}\right],
    \label{eqn:error-tr}
\end{align}
where $(\cdot)$ is either left or right. We compare the mean relative error $\Delta_{\text{left}}$($\Delta_{\text{right}}$) for NLCE with the corresponding mean relative error of the EE as a function of the number of zeros $m$ included in the best-fit line in Fig.~\ref{fig:tr_phase-boundary-err}. The extrapolated NLCE phase boundary from the right \textit{leg} in Fig.~\ref{fig:tr_phase-boundary-err}(b) has a much smaller error compared to EE, and it is robust $(\Delta_{\text{right}}\lesssim 5\%)$ for $1\le m \le 50$. The origin of this robustness is similar to what we explained in the square lattice, as the NLCE provides many more zeros before including the zeros close to the line $\Im(\beta h)=\pi/2$. For the left \textit{leg} $\mathcal{D}_{\text{left}}$, both NLCE and EE are not doing well in extrapolating the HKKK phase boundary, which is indicated by the high values of $\Delta_{\text{left}}$ in Fig.~\ref{fig:tr_phase-boundary-err}(a). The mismatch with the HKKK values could be attributed to a few reasons: (1) the choice of a best-fit line extrapolation was picked based on the results of the square lattice and the right \textit{leg} in the triangular lattice, which may not be the best choice for the left \textit{leg}. (2) The left \textit{leg} could be more prone to finite size effects compared to the right \textit{leg} as we saw in the convergence to the thermodynamic limit in Fig.~\ref{fig:tr_conv}. (3) To our best knowledge, the small-$T$ and small-$h$ part of the phase diagram has been harder to capture using numerical methods in literature~\cite{noh1992phase,hwang2008thermodynamic}.

It is also interesting to note the uncanny resemblance between the distribution of the zeros in the square lattice \textit{leg} $\mathcal{D}$ in Fig. \ref{fig:AFM_sq_h-plane} and the triangular lattice right \textit{leg} $\mathcal{D}_{\text{right}}$ in Fig.~\ref{fig:AFM_tr_h-plane}, which is supported by the similarities between $\alpha$ in Fig. \ref{fig:sq_conv} and $\alpha_{\text{right}}$ in Fig.~\ref{fig:tr_conv}(b) in addition to the small mean relative errors in the extrapolated phase boundary $\Delta$ in Fig.~\ref{fig:sq_phase-boundary}(b) and $\Delta_{\text{right}}$ in Fig.~\ref{fig:tr_phase-boundary-err}(b). All these similarities suggest that the nature of the phase transition upon increasing the magnetic field at a fixed $T$ in the square lattice is the same as the corresponding one in the triangular lattice at large $h$ (right \textit{leg}). This is also supported by the fact that these two transitions are transitions from an antiferromagnetic ordered phase to a paramagnetic phase, which is in contrast to the triangular lattice's left \textit{leg}, where the transition is between frustrated and antiferromagnetic states.

\section{Summary and Discussion}\label{Sec:Summary}
Our work gives more insights into the Yang-Lee zeros distribution in the thermodynamic limit of the Ising model with nearest-neighbor antiferromagnetic interactions for both the square and triangular lattices. We achieved this by implementing a numerical linked cluster expansion (NLCE) on the Yang-Lee zeros of finite clusters taking advantage of the free energy approach. For the square lattice, our results suggest that for $T\ll T_c$, the root curve is similar to the ferromagnetic case below $T_c$. Namely, the root curves are vertical lines in the complex $h$ plane, which are circles in the complex $z$ plane, that intersects the real axis at the critical fields. More analysis is still needed to explore the root curves for $T\lesssim T_c$, although our linear fits for the roots give reasonable agreement with the phase diagram. We also showed that using NLCE for root curve estimation and critical field extrapolation yields more robust and accurate results than exact enumeration (EE) since the zeros are much less sparse and zeros from rectangular-shaped clusters are included. 

For the triangular lattice, since the ordered phase has two sides, i.e. phase boundaries, we distinguished between the root curves that give rise to each phase boundary. The right boundary --the one at higher fields in magnitude-- gives root curves that are similar to the case of a square lattice. We believe this occurs since these two transitions are similar in nature as the system transitions from an antiferromagnetic ordered state to a paramagnetic state. However, the left boundary --the one at lower fields in magnitude-- gives less accurate results, albeit with reasonable agreement for most temperatures considered. This has been observed for other numerical techniques that attempted to find the phase boundary on this side as well. An important direction for future work is to better understand the distribution of zeros in this region of the complex $h$ plane, in particular which zeros survive the thermodynamic limit.

Our work introduces the utilization of NLCEs as a viable tool to study phase transitions through the Yang-Lee zeros approach. This can be further extended for other classical and quantum models in future studies. Also, the transition of the root curves from ferromagnetic to antiferromagnetic interactions for a given model remains an open question for future work. Finally, further exploration of the role the zeros' amplitudes play when considering the thermodynamic limit is an interesting question that we aim to study in future research. 

\acknowledgments
This work was supported by the National Science Foundation (NSF) Grant No.~PHY-2309146 (M.A.). M.A.~is grateful to Marcos Rigol and M.S.~to Sriram Shastry for stimulating discussions.


\bibliography{References}

\begin{thebibliography}{10}

\bibitem{pathria2011}
R.~K. Pathria and P.~D. Beale, {\em Statistical Mechanics}.
\newblock Amsterdam ; Boston: Elsevier/Academic Press, 3rd ed~ed., 2011.

\bibitem{domb_green_book_74}
C.~Domb and M.~S. Green, {\em Phase Transitions and Critical Phenomena}.
\newblock New York: Academic Press, 1974.

\bibitem{yang1952statistical}
C.-N. Yang and T.-D. Lee, ``Statistical theory of equations of state and phase transitions. i. theory of condensation,'' {\em Physical Review}, vol.~87, no.~3, p.~404, 1952.

\bibitem{van1984location}
W.~van Saarloos and D.~A. Kurtze, ``Location of zeros in the complex temperature plane: Absence of lee-yang theorem,'' {\em Journal of Physics A: Mathematical and General}, vol.~17, no.~6, p.~1301, 1984.

\bibitem{mussardo2017yang}
G.~Mussardo, R.~Bonsignori, and A.~Trombettoni, ``Yang--lee zeros of the yang--lee model,'' {\em Journal of Physics A: Mathematical and Theoretical}, vol.~50, no.~48, p.~484003, 2017.

\bibitem{bencs2022lee}
F.~Bencs, P.~Buys, L.~Guerini, and H.~Peters, ``Lee-yang zeros of the antiferromagnetic ising model,'' {\em Ergodic theory and dynamical systems}, vol.~42, no.~7, pp.~2172--2206, 2022.

\bibitem{lee1952statistical}
T.-D. Lee and C.-N. Yang, ``Statistical theory of equations of state and phase transitions. ii. lattice gas and ising model,'' {\em Physical Review}, vol.~87, no.~3, p.~410, 1952.

\bibitem{hemmer1964yang}
P.~Hemmer and E.~H. Hauge, ``Yang-lee distribution of zeros for a van der waals gas,'' {\em Physical Review}, vol.~133, no.~4A, p.~A1010, 1964.

\bibitem{sedik2024yang}
M.~Sedik, J.~M. Bhat, A.~Dhar, and B.~S. Shastry, ``Yang-lee zeros of certain antiferromagnetic models,'' {\em Physical Review E}, vol.~110, no.~1, p.~014117, 2024.

\bibitem{katsura1954phase}
S.~Katsura, ``On the phase transition,'' {\em The Journal of Chemical Physics}, vol.~22, no.~7, pp.~1277--1278, 1954.

\bibitem{heilmann1971location}
O.~J. Heilmann, ``Location of the zeros of the grand partition function of certain classes of lattice gases,'' {\em Studies in Applied Mathematics}, vol.~50, no.~4, pp.~385--390, 1971.

\bibitem{newman1974zeros}
C.~M. Newman, ``Zeros of the partition function for generalized ising systems,'' {\em Communications on Pure and Applied Mathematics}, vol.~27, no.~2, pp.~143--159, 1974.

\bibitem{lieb1981general}
E.~H. Lieb and A.~D. Sokal, ``A general lee-yang theorem for one-component and multicomponent ferromagnets,'' {\em Communications in Mathematical Physics}, vol.~80, no.~2, pp.~153--179, 1981.

\bibitem{biskup2000general}
M.~Biskup, C.~Borgs, J.~T. Chayes, L.~J. Kleinwaks, and R.~Koteck{\`y}, ``General theory of lee-yang zeros in models with first-order phase transitions,'' {\em Physical Review Letters}, vol.~84, no.~21, p.~4794, 2000.

\bibitem{deger2020lee}
A.~Deger, F.~Brange, and C.~Flindt, ``Lee-yang theory, high cumulants, and large-deviation statistics of the magnetization in the ising model,'' {\em Physical Review B}, vol.~102, no.~17, p.~174418, 2020.

\bibitem{li2023lee}
C.~Li and F.~Yang, ``Lee-yang zeros in the rydberg atoms,'' {\em Frontiers of Physics}, vol.~18, no.~2, p.~22301, 2023.

\bibitem{asano1970theorems}
T.~Asano, ``Theorems on the partition functions of the heisenberg ferromagnets,'' {\em Journal of the Physical Society of Japan}, vol.~29, no.~2, pp.~350--359, 1970.

\bibitem{vecsei2022lee}
P.~M. Vecsei, J.~L. Lado, and C.~Flindt, ``Lee-yang theory of the two-dimensional quantum ising model,'' {\em Physical Review B}, vol.~106, no.~5, p.~054402, 2022.

\bibitem{li2023yang}
H.~Li, X.-H. Yu, M.~Nakagawa, and M.~Ueda, ``Yang-lee zeros, semicircle theorem, and nonunitary criticality in bardeen-cooper-schrieffer superconductivity,'' {\em Physical Review Letters}, vol.~131, no.~21, p.~216001, 2023.

\bibitem{li2024yang}
H.~Li, ``Yang-lee zeros in quantum phase transitions: An entanglement perspective,'' {\em Phys. Rev. B}, vol.~111, p.~045139, Jan 2025.

\bibitem{bena2005statistical}
I.~Bena, M.~Droz, and A.~Lipowski, ``Statistical mechanics of equilibrium and nonequilibrium phase transitions: the yang--lee formalism,'' {\em International Journal of Modern Physics B}, vol.~19, no.~29, pp.~4269--4329, 2005.

\bibitem{peng2015experimental}
X.~Peng, H.~Zhou, B.-B. Wei, J.~Cui, J.~Du, and R.-B. Liu, ``Experimental observation of lee-yang zeros,'' {\em Physical review letters}, vol.~114, no.~1, p.~010601, 2015.

\bibitem{brandner2017experimental}
K.~Brandner, V.~F. Maisi, J.~P. Pekola, J.~P. Garrahan, and C.~Flindt, ``Experimental determination of dynamical lee-yang zeros,'' {\em Physical review letters}, vol.~118, no.~18, p.~180601, 2017.

\bibitem{gao2024experimental}
H.~Gao, K.~Wang, L.~Xiao, M.~Nakagawa, N.~Matsumoto, D.~Qu, H.~Lin, M.~Ueda, and P.~Xue, ``Experimental observation of the yang-lee quantum criticality in open quantum systems,'' {\em Physical Review Letters}, vol.~132, no.~17, p.~176601, 2024.

\bibitem{brange2024lee}
F.~Brange, N.~Lambert, F.~Nori, and C.~Flindt, ``Lee-yang theory of the superradiant phase transition in the open dicke model,'' {\em Physical Review Research}, vol.~6, no.~3, p.~033181, 2024.

\bibitem{heilmann1970monomers}
O.~J. Heilmann and E.~H. Lieb, ``Monomers and dimers,'' {\em Physical Review Letters}, vol.~24, no.~25, p.~1412, 1970.

\bibitem{lieb1972property}
E.~H. Lieb and D.~Ruelle, ``A property of zeros of the partition function for ising spin systems,'' {\em Journal of Mathematical Physics}, vol.~13, no.~5, pp.~781--784, 1972.

\bibitem{lebowitz2012location}
J.~L. Lebowitz, D.~Ruelle, and E.~R. Speer, ``Location of the lee-yang zeros and absence of phase transitions in some ising spin systems,'' {\em Journal of mathematical physics}, vol.~53, no.~9, p.~095211, 2012.

\bibitem{kim2004yang}
S.-Y. Kim, ``Yang-lee zeros of the antiferromagnetic ising model,'' {\em Physical review letters}, vol.~93, no.~13, p.~130604, 2004.

\bibitem{hwang2010yang}
C.-O. Hwang and S.-Y. Kim, ``Yang--lee zeros of triangular ising antiferromagnets,'' {\em Physica A: Statistical Mechanics and its Applications}, vol.~389, no.~24, pp.~5650--5654, 2010.

\bibitem{rigol_bryant_06}
M.~Rigol, T.~Bryant, and R.~R.~P. Singh, ``Numerical linked-cluster approach to quantum lattice models,'' {\em Phys. Rev. Lett.}, vol.~97, p.~187202, Nov 2006.

\bibitem{rigol_bryant_07a}
M.~Rigol, T.~Bryant, and R.~R.~P. Singh, ``Numerical linked-cluster algorithms. {I. S}pin systems on square, triangular, and kagom\'e lattices,'' {\em Phys. Rev. E}, vol.~75, p.~061118, Jun 2007.

\bibitem{rigol_bryant_07b}
M.~Rigol, T.~Bryant, and R.~R.~P. Singh, ``Numerical linked-cluster algorithms. {II. $t\text{\ensuremath{-}}J$} models on the square lattice,'' {\em Phys. Rev. E}, vol.~75, p.~061119, Jun 2007.

\bibitem{Tang_2013}
B.~Tang, E.~Khatami, and M.~Rigol, ``A short introduction to numerical linked-cluster expansions,'' {\em Computer Physics Communications}, vol.~184, pp.~557--564, mar 2013.

\bibitem{richter_20}
J.~Richter, T.~Heitmann, and R.~Steinigeweg, ``{Quantum quench dynamics in the transverse-field {I}sing model: A numerical expansion in linked rectangular clusters},'' {\em SciPost Phys.}, vol.~9, p.~031, 2020.

\bibitem{Abdelshafy_23}
M.~Abdelshafy and M.~Rigol, ``$\mathsf{L}$-based numerical linked cluster expansion for square lattice models,'' {\em Phys. Rev. E}, vol.~108, p.~034126, Sep 2023.

\bibitem{Abdelshafy_24}
M.~Abdelshafy and M.~Rigol, ``Numerical linked-cluster expansions for two-dimensional spin models with continuous disorder distributions,'' {\em Phys. Rev. E}, vol.~109, p.~054127, May 2024.

\bibitem{iyer_15}
D.~Iyer, M.~Srednicki, and M.~Rigol, ``Optimization of finite-size errors in finite-temperature calculations of unordered phases,'' {\em Phys. Rev. E}, vol.~91, p.~062142, Jun 2015.

\bibitem{bruognolo2017matrix}
B.~Bruognolo, Z.~Zhu, S.~R. White, and E.~M. Stoudenmire, ``Matrix product state techniques for two-dimensional systems at finite temperature,'' {\em arXiv preprint arXiv:1705.05578}, 2017.

\bibitem{khatami2014linked}
E.~Khatami, E.~Perepelitsky, M.~Rigol, and B.~S. Shastry, ``Linked-cluster expansion for the green's function of the infinite-u hubbard model,'' {\em Physical Review E}, vol.~89, no.~6, p.~063301, 2014.

\bibitem{hagymasi2021possible}
I.~Hagym{\'a}si, R.~Sch{\"a}fer, R.~Moessner, and D.~J. Luitz, ``Possible inversion symmetry breaking in the s= 1/2 pyrochlore heisenberg magnet,'' {\em Physical Review Letters}, vol.~126, no.~11, p.~117204, 2021.

\bibitem{biella2018linked}
A.~Biella, J.~Jin, O.~Viyuela, C.~Ciuti, R.~Fazio, and D.~Rossini, ``Linked cluster expansions for open quantum systems on a lattice,'' {\em Physical Review B}, vol.~97, no.~3, p.~035103, 2018.

\bibitem{richter2019combining}
J.~Richter and R.~Steinigeweg, ``Combining dynamical quantum typicality and numerical linked cluster expansions,'' {\em Physical Review B}, vol.~99, no.~9, p.~094419, 2019.

\bibitem{gan_20}
J.~Gan and K.~R.~A. Hazzard, ``Numerical linked cluster expansions for inhomogeneous systems,'' {\em Phys. Rev. A}, vol.~102, p.~013318, Jul 2020.

\bibitem{bhanot1990numerical}
G.~Bhanot, ``A numerical method to compute exactly the partition function with application to z (n) theories in two dimensions,'' {\em Journal of statistical physics}, vol.~60, no.~1-2, pp.~55--75, 1990.

\bibitem{creswick1995transfer}
R.~Creswick, ``Transfer matrix for the restricted canonical and microcanonical ensembles,'' {\em Physical Review E}, vol.~52, no.~6, p.~R5735, 1995.

\bibitem{Onsager_1944}
L.~Onsager, ``Crystal statistics. {I. A} two-dimensional model with an order-disorder transition,'' {\em Phys. Rev.}, vol.~65, pp.~117--149, Feb 1944.

\bibitem{wang1997critical}
X.-Z. Wang and J.~S. Kim, ``The critical line of an ising antiferromagnet on square and honeycomb lattices,'' {\em Physical review letters}, vol.~78, no.~3, p.~413, 1997.

\bibitem{muller1977interface}
E.~M{\"u}ller-Hartmann and J.~Zittartz, ``Interface free energy and transition temperature of the square-lattice ising antiferromagnet at finite magnetic field,'' {\em Zeitschrift f{\"u}r Physik B Condensed Matter}, vol.~27, no.~3, pp.~261--266, 1977.

\bibitem{monroe2001systematic}
J.~L. Monroe, ``Systematic approximation method for the critical properties of lattice spin systems,'' {\em Physical Review E}, vol.~64, no.~1, p.~016126, 2001.

\bibitem{shore2015charge}
J.~D. Shore and G.~M. Thurston, ``Charge-regulation phase transition on surface lattices of titratable sites adjacent to electrolyte solutions: An analog of the ising antiferromagnet in a magnetic field,'' {\em Physical Review E}, vol.~92, no.~6, p.~062123, 2015.

\bibitem{Wannier_1950}
G.~H. Wannier, ``Antiferromagnetism. {T}he triangular {I}sing net,'' {\em Phys. Rev.}, vol.~79, pp.~357--364, Jul 1950.

\bibitem{HOUTAPPEL1950425}
R.~Houtappel, ``Order-disorder in hexagonal lattices,'' {\em Physica}, vol.~16, no.~5, pp.~425--455, 1950.

\bibitem{schick1977phase}
M.~Schick, J.~Walker, and M.~Wortis, ``Phase diagram of the triangular ising model: Renormalization-group calculation with application to adsorbed monolayers,'' {\em Physical Review B}, vol.~16, no.~5, p.~2205, 1977.

\bibitem{noh1992phase}
J.~D. Noh and D.~Kim, ``Phase boundary and universality of the triangular lattice antiferromagnetic ising model,'' {\em International Journal of Modern Physics B}, vol.~6, no.~17, pp.~2913--2924, 1992.

\bibitem{hwang2008thermodynamic}
C.~Hwang, S.~Kim, D.~Kang, and J.~M. Kim, ``Thermodynamic properties of the triangular-lattice ising antiferromagnet in a uniform magnetic field,'' {\em JOURNAL-KOREAN PHYSICAL SOCIETY}, vol.~52, p.~S203, 2008.

\bibitem{stephenson1964ising}
J.~Stephenson, ``Ising-model spin correlations on the triangular lattice,'' {\em Journal of Mathematical Physics}, vol.~5, no.~8, pp.~1009--1024, 1964.

\bibitem{jacobsen1997monte}
J.~L. Jacobsen and H.~C. Fogedby, ``Monte carlo study of correlations near the ground state of the triangular antiferromagnetic ising model,'' {\em Physica A: Statistical Mechanics and its Applications}, vol.~246, no.~3-4, pp.~563--575, 1997.

\bibitem{kim2015ising}
S.-Y. Kim, ``Ising antiferromagnet on a finite triangular lattice with free boundary conditions,'' {\em Journal of the Korean Physical Society}, vol.~67, pp.~1517--1523, 2015.

\bibitem{monroe1998frustrated}
J.~L. Monroe, ``Frustrated ising systems on husimi trees,'' {\em Physica A: Statistical Mechanics and its Applications}, vol.~256, no.~1-2, pp.~217--228, 1998.

\end{thebibliography}
\bibliographystyle{ieeetr}

\end{document}